\documentclass[preprint2,iop,numberedappendix]{emulateapj}

\usepackage[caption=false]{subfig}
\usepackage{amsmath}
\usepackage{footnote}
\bibpunct{(}{)}{;}{a}{}{,} 
\newcommand{\dif}{\mathrm{d}}

\shorttitle{Detecting EoR in redshifted 21-cm}
\shortauthors{Thyagarajan et~al.}

\def\RRI{\altaffilmark{1}}
\def\RRItxt{\altaffiltext{1}{Raman Research Institute, Bangalore, India; e-mail: nithya.rri@gmail.com}}

\def\NRAONM{\altaffilmark{2}}
\def\NRAONMtxt{\altaffiltext{2}{National Radio Astronomy Observatory, Socorro, NM, USA}}

\def\CAASTRO{\altaffilmark{3}}
\def\CAASTROtxt{\altaffiltext{3}{ARC Centre of Excellence for All-sky Astrophysics (CAASTRO), Sydney, Australia}}

\def\Curtin{\altaffilmark{4}}
\def\Curtintxt{\altaffiltext{4}{Curtin University, Perth, Australia}}

\def\CfA{\altaffilmark{5}}
\def\CfAtxt{\altaffiltext{5}{Harvard-Smithsonian Center for Astrophysics, Cambridge, USA}}

\def\ASU{\altaffilmark{6}}
\def\ASUtxt{\altaffiltext{6}{Arizona State University, Tempe, USA}}

\def\ANU{\altaffilmark{7}}
\def\ANUtxt{\altaffiltext{7}{The Australian National University, Canberra, Australia}}

\def\CSIRO{\altaffilmark{8}}
\def\CSIROtxt{\altaffiltext{8}{CSIRO Astronomy and Space Science, NSW, Australia}}

\def\Haystack{\altaffilmark{9}}
\def\Haystacktxt{\altaffiltext{9}{MIT Haystack Observatory, Westford, USA}}

\def\USydney{\altaffilmark{10}}
\def\USydneytxt{\altaffiltext{10}{Sydney Institute for Astronomy, School of Physics, The University of Sydney, Sydney, Australia}}

\def\MIT{\altaffilmark{11}}
\def\MITtxt{\altaffiltext{11}{MIT Kavli Institute for Astrophysics and Space Research, Cambridge, USA}}

\def\UW{\altaffilmark{12}}
\def\UWtxt{\altaffiltext{12}{University of Washington, Seattle, USA}}

\def\Victoria{\altaffilmark{13}}
\def\Victoriatxt{\altaffiltext{13}{Victoria University of Wellington, Wellington, New Zealand}}

\def\UWisc{\altaffilmark{14}}
\def\UWisctxt{\altaffiltext{14}{University of Wisconsin--Milwaukee, Milwaukee, USA}}

\def\UMelbourne{\altaffilmark{15}}
\def\UMelbournetxt{\altaffiltext{15}{The University of Melbourne, Melbourne, Australia}}

\def\Tata{\altaffilmark{16}}
\def\Tatatxt{\altaffiltext{16}{Tata Institute of Fundamental Research, Pune, India}}

\def\NRAOGBT{\altaffilmark{17}}
\def\NRAOGBTtxt{\altaffiltext{17}{National Radio Astronomy Observatory, Charlottesville, USA}}

\def\UTasmania{\altaffilmark{18}}
\def\UTasmaniatxt{\altaffiltext{18}{University of Tasmania, Hobart, Australia}}

\begin{document}

\title{A study of fundamental limitations to statistical detection of redshifted HI from the epoch of reionization}



\author{
Nithyanandan~Thyagarajan\RRI,
N.~Udaya~Shankar\RRI,
Ravi~Subrahmanyan\RRI$^,$\NRAONM$^,$\CAASTRO,
Wayne~Arcus\Curtin, 
Gianni~Bernardi\CfA, 
Judd~D.~Bowman\ASU, 
Frank~Briggs\ANU$^,$\CAASTRO,
John~D.~Bunton\CSIRO, 
Roger~J.~Cappallo\Haystack, 
Brian~E.~Corey\Haystack, 
Ludi~deSouza\CSIRO$^,$\USydney,
David~Emrich\Curtin,
Bryan~M.~Gaensler\USydney$^,$\CAASTRO, 
Robert~F.~Goeke\MIT,
Lincoln~J.~Greenhill\CfA,
Bryna~J.~Hazelton\UW,
David~Herne\Curtin, 
Jacqueline~N.~Hewitt\MIT, 
Melanie~Johnston-Hollitt\Victoria,
David~L.~Kaplan\UWisc, 
Justin~C.~Kasper\CfA, 
Barton~B.~Kincaid\Haystack, 
Ronald~Koenig\CSIRO, 
Eric~Kratzenberg\Haystack, 
Colin~J.~Lonsdale\Haystack, 
Mervyn~J.~Lynch\Curtin, 
S.~Russell~McWhirter\Haystack,
Daniel~A.~Mitchell\UMelbourne$^,$\CAASTRO, 
Miguel~F.~Morales\UW, 
Edward~H.~Morgan\MIT, 
Divya~Oberoi\Tata, 
Stephen~M.~Ord\Curtin$^,$\CfA, 
Joseph~Pathikulangara\CSIRO, 
Ronald~A.~Remillard\MIT, 
Alan~E.~E.~Rogers\Haystack, 
D.~Anish~Roshi\NRAOGBT, 
Joseph~E.~Salah\Haystack, 
Robert~J.~Sault\UMelbourne, 
K.~S.~Srivani\RRI, 
Jamie~B.~Stevens\CSIRO$^,$\UTasmania, 
Prabu~Thiagaraj\RRI, 
Steven~J.~Tingay\Curtin$^,$\CAASTRO, 
Randall~B.~Wayth\Curtin$^,$\CfA$^,$\CAASTRO, 
Mark~Waterson\Curtin$^,$\ANU,
Rachel~L.~Webster\UMelbourne$^,$\CAASTRO, 
Alan~R.~Whitney\Haystack, 
Andrew~J.~Williams\Curtin, 
Christopher~L.~Williams\MIT, 
J.~Stuart~B.~Wyithe\UMelbourne$^,$\CAASTRO
}


\RRItxt
\NRAONMtxt
\CAASTROtxt
\Curtintxt
\CfAtxt
\ASUtxt
\ANUtxt
\CSIROtxt
\Haystacktxt
\USydneytxt
\MITtxt
\UWtxt
\Victoriatxt
\UWisctxt
\Tatatxt
\NRAOGBTtxt
\UMelbournetxt
\UTasmaniatxt



\begin{abstract}

In this paper we explore for the first time the relative magnitudes of three fundamental sources of uncertainty, namely, foreground contamination, thermal noise and sample variance in detecting the H{\sc i} power spectrum from the Epoch of Reionization (EoR). We derive limits on the sensitivity of a Fourier synthesis telescope to detect EoR based on its array configuration and a statistical representation of images made by the instrument. We use the Murchison Widefield Array (MWA) configuration for our studies. Using a unified framework for estimating signal and noise components in the H{\sc i} power spectrum, we derive an expression for and estimate the contamination from extragalactic point--like sources in three--dimensional $k$--space. Sensitivity for EoR H{\sc i} power spectrum detection is estimated for different observing modes with MWA. With 1000~hours of observing on a single field using the 128--tile MWA, EoR detection is feasible (S/N~$>1$ for $k\lesssim 0.8$~Mpc$^{-1}$). Bandpass shaping and refinements to the {\it EoR window} are found to be effective in containing foreground contamination, which makes the instrument tolerant to imaging errors. We find that for a given observing time, observing many independent fields of view does not offer an advantage over a single field observation when thermal noise dominates over other uncertainties in the derived power spectrum. 

\end{abstract}


\section{Introduction}\label{intro}

Precise measurements of the cosmic microwave background (CMB) anisotropies have constrained the background cosmology and initial conditions for structure formation. However, understanding the non-linear growth of density perturbations and astrophysical evolution in the Epoch of Reionization (EoR) has been difficult. Evidence to date suggests a complex reionization history  \citep{hai03,cen03,sok03,mad04};  for instance, the CMB data, when fitted to models of instantaneous reionization, point to a reionization redshift of $z\approx 10.5-11$ \citep{kog03,jar11,kom11,lar11}, which is in conflict with observations of Gunn-Peterson absorption troughs and near-zone transmission towards distant quasars indicating rapid evolution in the ionization fraction as late as $z\approx$ 6--7 \citep{bec01,djo01,fan02,mor11}. 

Direct observation of redshifted 21~cm spin transition of neutral hydrogen has been identified to be a useful method for detecting structures in cosmological gas at high redshifts \citep{sun72,sco90,mad97,toz00,ili02}. Tomography of redshifted 21~cm line promises to be a key probe of reionization history \citep{zal04}. Observing images of the three--dimensional distribution of neutral hydrogen temperature fluctuations in excess relative to the CMB temperature is expected to reveal the epoch as well as the process of reionization in detail; however, \citet{fur04} point out that such imaging requires the sensitivity of the Square Kilometer Array (SKA). Recently, through the use of simulations, the potential of SKA precursors for direct imaging and detection of ionized regions during late stages of reionization has been demonstrated \citep{zar12,mal13}. Numerous first-generation radio telescopes such as the Murchison Widefield Array \citep[MWA;][]{lon09,tin13}, the Low Frequency Array \citep[LOFAR;][]{van13}, and the Precision Array for Probing the Epoch of Reionization \citep[PAPER;][]{par10} are becoming operational with enough sensitivity for a statistical detection of the EoR H{\sc i} power spectrum. Measuring the H{\sc i} power spectrum and its cosmological evolution is a first step to understanding structure formation and astrophysics in the EoR.

Power spectrum measurements of the redshifted 21~cm from EoR are difficult for the following reasons. The EoR signal is extremely weak relative to the foreground emission of the Galaxy and extragalactic sources \citep{ber09,gho12}. Considerable effort is required to distinguish their signatures from residual errors even after careful spectral modeling and subtraction of these foregrounds \citep{dim02,zal04}. \citet{mor04} show that the inherent isotropy and symmetry of the EoR signal in frequency and spatial wavenumber ($k$) space make it distinguishable from sources of contamination which lack such symmetry. But they note that such symmetry considerations provide only an additional tool for separating foreground contamination from the signal, and do not guarantee that foreground contamination will be removed. 

An inherent mechanism of foreground contamination via the frequency dependent structure (chromaticity) of the primary and synthesized beams has been pointed out by \citet{bow09} and \citet{mor12}. The chromatic nature of the primary and synthesized beams carries the transverse structure of contamination due to the residuals of continuum foreground subtraction into the line-of-sight direction. This has been termed {\it mode--mixing}. Both analytic calculations of \citet{ved12} and simulations of \citet{dat10} and \citet{tro12} have shown that foreground contamination by residuals after source subtraction are predominantly localized to a wedge-like region in $k$--space. The region excluded by the wedge has been termed as the {\it EoR window} \citep{mor12,ved12}. They have also indicated that appropriate choices of bandpass window functions and imaging algorithms can significantly minimize levels of such contamination in specific regions of $k$--space. 

In this paper, we present a unified framework for estimating three fundamental sources of uncertainty, namely, foreground contamination, thermal noise and sample variance in $k$--space. We apply this general understanding to the case of MWA using different observing modes. We have also explored the effects of shaping the bandpass window and refining the {\it EoR window}. With detailed estimates, we compare the relative magnitudes of different sources of uncertainties and obtain a more complete view of the EoR sensitivity of the 128--tile MWA. 

The rest of the paper is organized as follows. \S\ref{sec:theory} provides a quick snapshot of the cosmology that motivates radio observations. \S\ref{sec:measurements} sets up the basic radio interferometer measurements of signal and uncertainties. Parameters and notations used are introduced that bridge the radio interferometer measurements and cosmological motivations. \S\ref{sec:framework} introduces the framework upon which we build our understanding and estimates of different sources of uncertainties. Here, we also list some assumptions that have gone into our study. In \S\ref{sec:foreground}, and \S\ref{sec:TN}, we describe the $k$--space occupancy and estimates of foreground and thermal noise components, respectively, in the power spectrum. \S\ref{sec:EoR_PS_SV} provides estimates of the EoR H{\sc i} power spectrum and sample variance. The detailed interplay between various uncertainties under different observing modes and instrument parameters in determining sensitivity of the instrument for statistical measurements of EoR signatures is discussed in \S\ref{sec:EoR_detection}. The results are then summarized in \S\ref{sec:summary}. In appendices \S\ref{app:confusion_theory} and \S\ref{app:FG}, we provide the details behind the derivation and estimation of classical radio source confusion and power spectrum of extragalactic point--like sources in $k$--space, respectively.

\section{Basic Theory}\label{sec:theory}

\citet{lid08} provide a basis for understanding the power spectrum of the 21~cm brightness temperature (relative to CMB) fluctuations in the limit that the spin temperature, $T_\textrm{S}$, is globally much larger than the CMB temperature, $T_\textrm{CMB}$. Ignoring peculiar velocities, the 21~cm brightness temperature relative to the CMB at spatial position, $\overline{\mathbf{r}}$, is, 
\begin{equation}
  \delta_T(\overline{\mathbf{r}})=T_0\,\langle x_\textrm{H}\rangle\,\big[1+\delta_x(\overline{\mathbf{r}})\big]\,\big[1+\delta_\rho(\overline{\mathbf{r}})\big].
\end{equation}
Here, $T_0$ is the 21~cm brightness temperature of a neutral gas element with cosmic mean gas density, at redshift $z$, observed at frequency $f_0=1420/\left(1+z\right)$~MHz, relative to the CMB temperature at that epoch. $T_{0}=28\,\big[\left(1+z\right)/10\big]^{1/2}$~mK for the cosmological parameters we have adopted throughout this paper (symbols have their usual meanings): $H_0=70$~km~s$^{-1}$~Mpc$^{-1}$, $\Omega_m=0.27$, $\Omega_\Lambda=0.73$, and $\Omega_K=1-\Omega_m-\Omega_\Lambda$. $\langle x_\textrm{H}\rangle$ is the volume-averaged neutral fraction, $\delta_x$ is the fractional fluctuation in the neutral fraction, and $\delta_{\rho }$ is the fractional gas density fluctuation. The volume-averaged ionization fraction is, $\langle x_i\rangle=1-\langle x_\textrm{H}\rangle$. 

The power spectrum of $\delta_T(\overline{\mathbf{r}})$ is given in $k$--space by $P^\textrm{H{\sc i}}(\overline{\mathbf{k}})$, which is the Fourier transform of $\bigl\langle\delta_T(\overline{\mathbf{r}})\,\delta_T(\overline{\mathbf{r}}+\Delta\overline{\mathbf{r}})\bigr\rangle$, and $\overline{\mathbf{k}}$ is the Fourier conjugate variable of $\Delta\overline{\mathbf{r}}$. Assuming isotropy of neutral hydrogen distribution, $P^\textrm{H{\sc i}}(\overline{\mathbf{k}})$ may be described using only the radial coordinate $k$, as $P^\textrm{H{\sc i}}(k)$. Equivalently, the dimensionless quantity $\Delta_\textrm{H{\sc i}}^2(k)$ is frequently used to represent power in a logarithmic interval of $k$, given by \citep{zal04,mcq06,lid08},
\begin{equation}\label{eqn:EoR_PS}
  \Delta_\textrm{H{\sc i}}^2(k)=k^3P^\textrm{H{\sc i}}(k)/2\pi^2T_0^2.
\end{equation} 

\section{Interferometer Measurements in the EoR Context}\label{sec:measurements}

$P^\textrm{H{\sc i}}(k)$ is estimated using the image cube, $I(l,m,f)$, representing the sky brightness distribution in ($l,m,f$)--coordinates. In radio interferometry, $I(l,m,f)$ is obtained by Fourier transforming the visibility measurements, $V(u,v,f)$, made in ($u,v,f$)--coordinates. $u$ and $v$ are baseline lengths in units of wavelength, and $l$ and $m$ denote the direction cosines on the celestial sphere. $f$ denotes the frequency of observation. $\eta$ represents instrumental delay. $(l,m,f)$ and $(u,v,\eta)$ form a Fourier conjugate pair of variables. We adopt the following convention for Fourier transform:
\begin{align}
  V(u,v,\eta) &= \iiint I(l,m,f) e^{-j2\pi(ul+vm+\eta f)}\,\dif l\,\dif m\,\dif f, \\
  I(l,m,f) &= \iiint V(u,v,\eta)\,e^{j2\pi(ul+vm+\eta f)}\,\dif u\,\dif v\,\dif \eta,
\end{align}
where $j=\sqrt{-1}$. 

The sky brightness distribution comprising of the true EoR H{\sc i} signal and foregrounds is multiplied by the primary beam power pattern, $W_{lm}^\textrm{P}(l,m)$, on the sky. Equivalently, the visibilities of true EoR H{\sc i} signal, $V_{uvf}^\textrm{H{\sc i};T}(u,v,f)$, and foregrounds, $V_{uvf}^\textrm{FG;T}(u,v,f)$, are convolved with the spatial frequency response of the power pattern of an individual antenna, $W_{uv}^\textrm{P}(u,v)$. $W_{lm}^\textrm{P}(l,m)$ and $W_{uv}^\textrm{P}(u,v)$ form a Fourier transform pair. The convolved visibilities are corrupted by additive thermal noise, $V_{uvf}^\textrm{N}(u,v,f)$, sampled at the baseline locations given by the sampling function, $S_{uv}(u,v)$. The sampling function is the Fourier counterpart of the synthesized beam, $S_{lm}(l,m)$. Along the line of sight, the visibilities are modified by the frequency bandpass weights, $W_f^\textrm{B}(f)$. Thus, the measured visibilities may be expressed as:
\begin{align}\label{eqn:obsvis}
V_{uvf}^\textrm{obs}(u,v,f) &= \Big\{\left[V_{uvf}^\textrm{H{\sc i};T}(u,v,f)+V_{uvf}^\textrm{FG;T}(u,v,f)\right] \notag\\
& \qquad\quad \ast W_{uv}^\textrm{P}(u,v) + V_{uvf}^\textrm{N}(u,v,f)\Big\}\notag\\
& \qquad\quad \,S_{uv}(u,v)\,W_f^\textrm{B}(f), 
\end{align}
where the symbol $\ast$ denotes convolution. Equation~(\ref{eqn:obsvis}) forms the basis of our estimates of signal and various components of uncertainty in the power spectrum. A matrix--based framework is described in \citet{liu11}.

In Fourier space, $V_{uvf}^\textrm{obs}(u,v,f)$, is transformed to:
\begin{align}\label{eqn:obsvis_fourier}
V_{uv\eta}^\textrm{obs}(u,v,\eta) &= \int V_{uvf}^\textrm{obs}(u,v,f)\,e^{-j2\pi\eta f}\dif f.
\end{align}
This is obtained by Fourier transforming along frequency. 

The characteristic size of the spatial frequency response of the tile's power pattern, $W_{uv}^\textrm{P}(u,v)$, is $A_\textrm{e}/\lambda^2$, where $A_\textrm{e}$ is the effective area of a tile and $\lambda$ is the observing wavelength. The Fourier response of bandpass window, $W_f^\textrm{B}(f)$, is $W_\eta^\textrm{B}(\eta)$. The characteristic width of the bandpass response function, $W_\eta^\textrm{B}(\eta)$, is set by the inverse of effective bandwidth, $B_\textrm{eff}$.
$W_{uv\eta}(u,v,\eta)=W_{uv\eta}^\textrm{P}(u,v,\eta)\ast W_\eta^\textrm{B}(\eta)$ is the instrumental response in the spatial frequency domain, where $W_{uv\eta}^\textrm{P}(u,v,\eta)$ may be interpreted as the spatial frequency response of the tile's power pattern obtained over a uniform and infinite bandpass window, and $W_{uv}^\textrm{P}(u,v)=W_{uv\eta}^\textrm{P}(u,v,\eta=0)$. Here, we have assumed that the power pattern of the tile does not vary significantly over the chosen frequency band. $W_{uv\eta}(u,v,\eta)$ is set such that $\int W_{uv\eta}\,\dif u\,\dif v\,\dif\eta = 1$, where $\delta u\,\delta v\,\delta\eta \simeq A_\textrm{e}/(\lambda^2 B_\textrm{eff})$. In our adopted Fourier convention, the primary beam, $W_{lm}^\textrm{P}(l,m)$, and bandpass window function, $W_f^\textrm{B}(f)$, each have peaks of value unity.

True sky visibilities are uncorrelated between non-identical baseline vectors (spatial frequencies). $W_{uv\eta}(u,v,\eta)$ is an instrumental function that introduces correlations between true sky visibilities. $W_{uv}^\textrm{P}(u,v)$ and $W_\eta^\textrm{B}(\eta)$ are the transverse and line-of-sight components, respectively, of this correlating instrumental function. $W_{uv}^\textrm{P}(u,v)$ is a convolution of the electric field distribution over the tile with itself which is, alternatively, the Fourier transform of the power pattern of the tile. If the electric field pattern of the tile was a uniform square, $W_{uv}^\textrm{P}(u,v)$ takes the shape of a square pyramid in the ($uv$)--plane. 

We use the array configuration of MWA to estimate the signal and uncertainties. For such an estimation, we gridded the ($uv$)--plane with a cell size of $\lambda/2$ on each side. This is to avoid aliasing effects in the image up to the spatial scale of the horizon, $(l,m)\in [-1,+1]$. $W_{uv}^\textrm{P}(u,v)$ is obtained on the grid by the convolution described above. We assume each MWA array element to be a 4$\times$4 array of identical radiators. This results in a discrete form of the square pyramid for $W_{uv}^\textrm{P}(u,v)$. Some authors \citep{mcq06,bow06,bow07,bea13} have assumed that $W_{uv\eta}(u,v,\eta)$ is a sharply peaked function, and hence approximated it by a {\it delta} function. In our study, we use the full functional form. It is important to keep this functional form in order to take into account the {\it multi--baseline mode--mixing}, which has been described in \citet{haz13}. The shape of $W_{uv}^\textrm{P}(u,v)$ we use is also consistent with the baseline power response shown in \citet{haz13}. 

We place the synthesized baselines represented by the sampling function $S_{uv}(u,v)$ as a two--dimensional histogram on this grid. Compared to an optimal gridding scheme, the histogram method could place the baselines in the grid with a maximum error of $\lambda/2$. This could cause a jitter in the weights on the grid which is most severe on scales comparable to the cell size in the grid (corresponds to spatial scale of horizon or larger). While such a jitter will be important in case of a synthesis imaging procedure, it is not as significant in determining grid weights in our study. 

In describing the aforementioned radio interferometer quantities and desired cosmological measurements, we use the following notation interchangeably throughout this paper:
\begin{equation}
  \underline{\mathbf{u}} \equiv (u,v), \,\, \underline{\mathbf{l}} \equiv (l,m), \,\, \overline{\mathbf{u}} \equiv (\underline{\mathbf{u}},\eta) \,\, \mathrm{and} \,\, \overline{\mathbf{l}} \equiv (\underline{\mathbf{l}},f).
\end{equation}
The spatial wave vectors, $\overline{\mathbf{k}}\equiv (\underline{\mathbf{k}},k_\parallel)\equiv (k_x,k_y,k_\parallel)$, are related to $\overline{\mathbf{u}}$ as \citep{mor04}:
\begin{align}
  \underline{\mathbf{k}} = \frac{2\pi\,\underline{\mathbf{u}}}{D(z)}, &\quad k_\parallel \approx \frac{2\pi\,\eta}{\left(\frac{c\,(1+z)^2}{H_0\,f_{21}\,E(z)}\right)}, \\
  \mathrm{with} \quad \Delta D(z) &\approx \frac{c\,(1+z)^2}{H_0\,f_{21}\,E(z)}\,\Delta f,
\end{align}
where $H_0$ and $E(z)\equiv [\Omega_\textrm{M}(1+z)^3+\Omega_k(1+z)^2+\Omega_\Lambda]^{1/2}$ are standard terms in cosmology and $D(z)$ is the transverse comoving distance at redshift $z$ \citep{hog99}. $f_{21}$ is the rest frequency of the 21~cm line. In the last equation, $\Delta D(z)$ may be identified as the line-of-sight comoving width of the observation at redshift $z$ if $\Delta f$ is set to the observing bandwidth. The following relations may also be noted: 
\begin{equation}\label{eqn:k-coords}
  k_\perp=\left|\:\underline{\mathbf{k}}\:\right|=(k_x^2+k_y^2)^{1/2}\quad \mathrm{and} \quad k=\left|\:\overline{\mathbf{k}}\:\right|=(k_\perp^2+k_\parallel^2)^{1/2}. 
\end{equation}
$k_\perp$ and $k_\parallel$ may be viewed as components of $\overline{\mathbf{k}}$ along the transverse and line-of-sight directions respectively. 

\begin{figure}[htb]
\centering
\includegraphics[width=\linewidth]{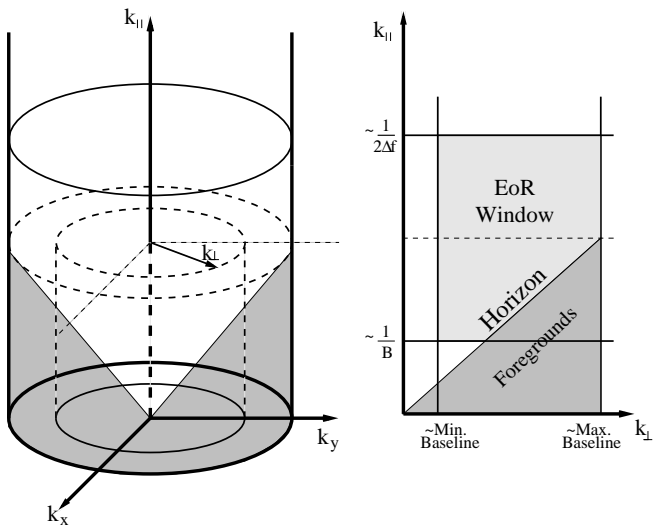}
\caption{Cartoon illustration of regions in $k$--space, including instrumental, foreground, and EoR windows. The panel on the left shows the three--dimensional $k$-- space. The panel on the right is obtained by collapsing $k$--space along $k_x$ and $k_y$ into $k_\perp$ using equation~(\ref{eqn:k-coords}). The shaded volume in the panel on the left is now reduced to a wedge-shaped region \citep{dat10,ved12} shown in a darker shade in the panel on the right. Being dominated by foreground sources and their sidelobes due to frequency dependent synthesized beams, this region is referred to as the ``foreground window''. The unshaded region excluded by the shaded cylindrical volume is referred to as the {\it EoR window}, where the signal is believed to be relatively free of such contaminations. This reduces to the region shown in a lighter shade in the panel on the right. The axes are not to scale. \label{fig:kspace}}
\end{figure}

Figure~\ref{fig:kspace} is a cartoon illustration of different regions of significance in $k$--space. The range in $k_\perp$ is set by the minimum and maximum baseline lengths while that along $k_\parallel$ is set by the channel resolution and bandwidth. This region is the instrumental window. Residuals from unsubtracted foregrounds and the structure of their frequency dependent sidelobes occupy the wedge-shaped region labeled as ``foregrounds''. The part of the instrumental window excluding the wedge, called the {\it EoR window}, is also shown. 
 
The radio measurements described above are related to the desired cosmological quantities. The power spectrum of EoR H{\sc i} fluctuations is related to the diagonal of the covariance matrix of true H{\sc i} visibilities in Fourier space as \citep{mor04,mor05}:
\begin{align}
  P^\textrm{H{\sc i}}(\overline{\mathbf{u}}) &= \Bigl\langle V_{uv\eta}^\textrm{H{\sc i};T}(\overline{\mathbf{u}}_i)^\star V_{uv\eta}^\textrm{H{\sc i};T}(\overline{\mathbf{u}}_j)\Bigr\rangle \; \delta_{ij} \notag\\
  &= \Bigl\langle \left|V_{uv\eta}^\textrm{H{\sc i};T}(\overline{\mathbf{u}})\right|^2\Bigr\rangle, \\
  \mathrm{where} \qquad P^\textrm{H{\sc i}}(\overline{\mathbf{u}}) &= P^\textrm{H{\sc i}}(\overline{\mathbf{k}})\,\left(\frac{1}{D}\right)^2\left(\frac{B_\textrm{eff}}{\Delta D}\right).\label{eqn:jacobian}
\end{align}
The last equation is obtained using the Fourier conventions and {\it Jacobian} in the transformation between quantities in ($k_x,k_y,k_\parallel$)-- and ($u,v,\eta$)--coordinates. 

\section{Framework}\label{sec:framework}

Following the notations established in \S\ref{sec:measurements}, we start with visibility measurements, $V(u,v,f)$, which when Fourier transformed along ($u,v$)--coordinates yield an image cube $I(l,m,f)$. From $I(l,m,f)$ we assume that the extragalactic point--like foreground sources have been perfectly removed down to a certain source confusion threshold, to obtain a residual image cube $\Delta I(l,m,f)$, which consists of unsubtracted sources and their sidelobes. A statistical representation of $\Delta I(l,m,f)$, together with the sampling in ($u,v$)--plane given by the array configuration, form the basis of our understanding of different uncertainties. 

\subsection{Instrument Properties and Adopted EoR Model}\label{sec:instr_parms}

Sampling of the ($u,v$)--plane is provided by the 128--tile MWA array configuration \citep{bea12}, in which the array elements (tiles) are quasi-randomly distributed with 112 of them distributed over an aperture 1.5~km in diameter and a small number (16) of outliers extending to 3~km. 

We have adopted a {\it natural} weighting scheme where each visibility measurement has equal weight. The weight of a ($uv$)--cell is proportional to the number of baselines, including redundant ones, that fall inside the cell. Such a weighting has a better sidelobe response and thermal noise sensitivity, while emphasizing short spacings (large scale structures) of the interferometer array compared to {\it uniform} weighting \citep{tay99,bow09}. 

We note that \citet{lid08} predict the amplitude of the EoR power spectrum to peak at $\langle x_i\rangle\sim 0.5$ and $z\sim 7.3$. This determined our choice of observing frequency, $f_0=170.7$~MHz, at which the system temperature is $T_\textrm{sys}\lesssim440$~K \citep{bow06} and the effective area of a tile is $A_\textrm{e}\simeq 12.3$~m$^2$ \citep{bow06}. The frequency resolution of MWA (40~kHz) is used. A bandwidth of 8~MHz is chosen to have minimal EoR signal evolution with redshift \citep{bow06,mcq06}. Some relevant instrument parameters are listed in Table \ref{tab:inst_parms}.

\begin{table}
\caption[]{Properties of 128--tile MWA used in our model observations.\label{tab:inst_parms}}
\centering
\begin{tabular}{lcc}
\tableline\tableline
Parameter & Symbol & Value \\
\tableline
Number of tiles & & 128 \\
Center frequency & $f_0$ & 170.7~MHz ($z\sim 7.3$) \\
Effective bandwidth & $B_\textrm{eff}$ & 8~MHz \\
Channel resolution & $\Delta f$ & 40~kHz \\
Effective area of tile & $A_\textrm{e}$ & 12.3~m$^2$ \\
System temperature & $T_\textrm{sys}$ & 440~K\\
Integration time & $t_\textrm{int}$ & 8~seconds\\
\tableline
\end{tabular}
\end{table}

For the model power spectrum $P^\textrm{H{\sc i}}(\overline{\mathbf{k}})$ we have chosen, the expected signal strength ranges from 1--10$^7$ (in observer's units of K$^2$~Hz$^2$) for $0.01$~Mpc$^{-1}\lesssim k\lesssim 4$~Mpc$^{-1}$. The strength of the signal expected in observations made with the 128--tile MWA is discussed in detail in \S\ref{sec:EoR_PS_SV}. 

\subsection{Case Studies of Model Observations}

We consider observations in which the MWA array is pointed at a declination $\delta=-26.7$\arcdeg. This is equal in value to the latitude of MWA and, hence, passes through zenith. We investigate two observing modes:
\begin{enumerate}
\item 6~hours synthesis on a single patch of sky repeated about 160 times to get a total observing time of 1000~hours; and, 
\item 6~hours synthesis on 20 different patches of sky (different RA with the same declination), each observed about 8 times (also amounting to a total observing time of 1000~hours), where each patch is separated from others by at least one FWHM of the primary beam\footnote{We model the primary beam power pattern of the MWA tile as a 4$\times $4 array of identical radiators. At 170.7~MHz, the primary beam has a FWHM of $\theta_\textrm{P}\approx 21$\arcdeg. A strip of 24~hour range in RA at a declination $\delta$ with a declination width $\Delta \delta$ subtends a solid angle $\Delta\Omega=4\pi\sin\frac{\Delta\delta}{2}\cos\delta$. For $\delta=-26.7$\arcdeg and $\Delta\delta=\theta_\textrm{P}$, the strip subtends a solid angle of $\approx$~2~sr and corresponds to $\approx$~20 non-overlapping primary beams at the specified FWHM.}. 
\end{enumerate}
The choice of 6~hours of aperture synthesis was made to confine the observations to within $\pm$3~hours in hour angle on either side of zenith as per the MWA EoR observing plan \citep{bea13}. The different observing modes used are summarized in Table~\ref{tab:obs_parms}. The first column refers to the numbering of different observational case studies, the second refers to the time of synthesis, the third refers to the number of independent patches of the sky observed, the fourth refers to the number of times each of these fields are observed with their respective times of synthesis, the fifth denotes the total amount of time spent observing each field and the sixth column lists the total observing time used in each case study. We assume $T_\textrm{sys}$ is identical for all patches of sky in these observing modes.

\begin{table}
\caption{Parameters used in different case studies of model observations.\label{tab:obs_parms}}
\centering
\begin{tabular}{cccccc}
\tableline\tableline
S.~No. & $t_\textrm{syn}$ & $N_\textrm{fields}$ & $N_\textrm{cad}$ & $t_\textrm{field}$ & $t_\textrm{obs}$ \\ 
\tableline
1 & 6~hours & 1 & 166.7 & 1000~hours & 1000~hours \\
2 & 6~hours & 20 & 8.3 & 50~hours & 1000~hours \\
\tableline
\end{tabular}
\end{table}

The motivation to explore model observations with multiple fields of view is as follows: if the total time spent observing a single field is divided over multiple fields, an independent measurement of the power spectrum for each of the fields will be obtained. Upon averaging these power spectra, different components of uncertainty, specifically sample variance, will be reduced. However, since the observing time on each field has reduced, the thermal noise component in individual power spectra obtained over each field will be worse than when all the time was spent observing a single field. This is because all visibilities in a single--field observation will be combined coherently before estimating the power spectrum unlike that in a multi--field observation. What are the relative levels of various components of uncertainty in the measured power spectrum? Are there regions in $k$--space where sample variance is the dominant source of uncertainty relative to the thermal noise component? And can sensitivity be improved in these regions by averaging power spectrum measurements from independent fields which reduces sample variance? Should a particular observing mode be preferred over others?  

\subsection{Assumptions}

We only consider point--like extragalactic sources in our analysis and leave the treatment of extended emission from the Galaxy and extragalactic sources to future work. Extragalactic point--like sources above the detection threshold and their sidelobes are assumed to be perfectly subtracted. For simplicity, we assume the flux densities of residual sources are constant with frequency over the band of interest after source subtraction. 

$T_\textrm{sys}$ could change at most by $\sim 6$\% over an 8~MHz band relative to the mean value at 170.7~MHz, assuming a synchrotron temperature spectral index of $\gamma=-2.5$ ($T\propto f^{\gamma}$). For this study, we assume $T_\textrm{sys}$ is constant over an 8~MHz frequency band. 

The bandpass can only be determined as accurately as the continuum model we have for the sky because it is solved for using the sky model including frequency structure of sources. Hence, errors in calibration of amplitude and phase that result in imaging errors will then lead to errors in deriving accurate bandpass calibration as well. The frequency structure of the bandpass could also be affected due to radio frequency interference (RFI). In this paper, we neglect effects of calibration errors and RFI. We also neglect the effects of non-coplanarity of baselines. 

\section{Foreground Power Spectrum}\label{sec:foreground}

One of the major contaminants in the EoR H{\sc i} signal is the synchrotron emission from extragalactic and Galactic foregrounds \citep{dim02,zal04}. Thus far, neither of these causes of foreground contamination have been accounted for in the estimates of sensitivity of three--dimensional EoR H{\sc i} power spectrum in a comprehensive manner. For instance, \citet{bea13} estimated sensitivity by excluding a wedge shaped region, thereby removing a majority of foreground contamination. But they did not consider possible spillover from this contamination. Does it imply with certainty that foregrounds play no role any more in contaminating the {\it EoR window} or in estimates of sensitivity? Detailed estimates of extragalactic foreground contamination as done below in this paper show that contamination is also present in the {\it EoR window} outside the wedge-shaped region, even after a perfect subtraction of foregrounds. 

\subsection{Classical Radio Source Confusion}\label{sec:clas_src_conf}

The foreground contamination in the sought EoR H{\sc i} power spectrum is seeded by classical radio source confusion. The cause of classical confusion and the theory behind it have been well studied in literature \citep{con74,roh00}. The basic ingredients for estimating classical confusion are radio source count statistics on the sky and instrument parameters such as synthesized beam size.  

Assuming sources brighter than five times the classical source confusion noise and their sidelobes have been subtracted perfectly, the source confusion variance, $\sigma_\textrm{C}^2(\underline{\mathbf{l}})\equiv P_{lm}^\textrm{FG}(\underline{\mathbf{l}})$, was estimated across the residual image, $\Delta I(l,m,f)$, using the source count statistics provided by \citet{hop03} at a frequency of 1.4~GHz. $\sigma_\textrm{C}^2$ increases with the solid angle subtended by an image pixel in ($l,m$)--coordinates, which in turn is a function of the pixel location. Hence, $\sigma_\textrm{C}^2$ varies with $\underline{\mathbf{l}}$. 

In obtaining the confusion variance at 170.7~MHz, we used a mean spectral index of $\alpha=-0.78$ \citep[$S\,\propto f^\alpha$]{ish10}. Flux densities are converted to temperature units using $T=S A_\textrm{e}/(2\,k_\textrm{B})$, where $k_\textrm{B}$ is the Boltzmann constant. Hereafter, fluctuations in the residual image cube are statistically represented by the classical source confusion variance $P_{lm}^\textrm{FG}(\underline{\mathbf{l}})$. 

In a naturally weighted image made using the 128--tile MWA, a typical pixel close to the zenith subtends a solid angle of $\Omega\approx 2\times 10^{-6}$~sr. Statistically, the flux contained in such pixels have uncertainties whose rms is $\sigma_\textrm{C}\approx 35$~mJy. 

\S\ref{app:confusion_theory} gives details of an iterative procedure used to arrive at the classical source confusion rms $\sigma_\textrm{C}$. An illustration of the numerical solution for the variation of source confusion variance with solid angle, which varies across the residual image, is also presented.

\subsection{Foregrounds in $k$--space}\label{sec:FG_kspace}

The foreground component of measured visibilities may be written from equation~(\ref{eqn:obsvis}) as:
\begin{align}
   V_{uvf}^\textrm{FG}(\underline{\mathbf{u}},f) &= V_{uvf}^\textrm{FG;T}(\underline{\mathbf{u}},f) \ast W_{uv}^\textrm{P}(\underline{\mathbf{u}})\,S_{uv}(\underline{\mathbf{u}})\,W_f^\textrm{B}(f).
\end{align}
Thus, 
\begin{align}\label{eqn:main:FG_vis_gen}
  V_{uvf}^\textrm{FG}(\overline{\mathbf{u}}) &= V_{uv\eta}^\textrm{FG;T}(\overline{\mathbf{u}})\,S_{uv}(\underline{\mathbf{u}}) \ast W_{uv}^\textrm{P}(\underline{\mathbf{u}}) \ast W_\eta^\textrm{B}(\eta),
\end{align}
where the true foreground visibilities, $V_{uvf}^\textrm{FG;T}(\underline{\mathbf{u}},f)$, have been convolved with the spatial frequency response of the tile's power pattern, $W_{uv}^\textrm{P}(\underline{\mathbf{u}})$, multiplied by the sampling function in the ($u,v$)--plane, $S_{uv}(\underline{\mathbf{u}})$, multiplied by the frequency bandpass window function, $W_f^\textrm{B}(f)$, and Fourier transformed along $f$.
 
The power spectrum of foregrounds is extracted from the diagonal of the covariance matrix $\Bigl\langle V_{uv\eta}^\textrm{FG}(\overline{\mathbf{u}}_i)^\star V_{uv\eta}^\textrm{FG}(\overline{\mathbf{u}}_j)\Bigr\rangle$. After certain algebraic simplifications between quantities in Fourier ($\overline{\mathbf{u}}$)-- and real ($\overline{\mathbf{l}}$)--space, the foreground component of power spectrum may be expressed as:
\begin{align}\label{eqn:main:PS_FG_mm}
  P^\textrm{FG}_\textrm{inst}(\overline{\mathbf{u}}) &= \iint P_{lm}^\textrm{FG}(\underline{\mathbf{l}})\,\ast\,\left|S_{lm}(\underline{\mathbf{l}})\right|^2\,\left|W_{lm}^\textrm{P}(\underline{\mathbf{l}})\right|^2\notag\\
  & \qquad \ast\,\left|W_\eta^\textrm{B}(\eta)\right|^2\,\ast\,\delta\left(\eta+\frac{\underline{\mathbf{x}}\cdot\underline{\mathbf{l}}}{c}\right)\,\dif^2 \underline{\mathbf{l}}, 
\end{align} 
where $P_{lm}^\textrm{FG}(\underline{\mathbf{l}})\equiv \sigma_\textrm{C}^2(\underline{\mathbf{l}})$. The classical source confusion variance estimated across the image is weighted by primary beam squared $\left|W_{lm}^\textrm{P}(\underline{\mathbf{l}})\right|^2$, then convolved with the synthesized beam squared $\left|S_{lm}(\underline{\mathbf{l}})\right|^2$, and is finally convolved with the instrumental delay response squared $\left|W_\eta^\textrm{B}(\eta)\right|^2$, which is shifted in $\eta$ by $\underline{\mathbf{x}}\cdot\underline{\mathbf{l}}/c$ as prescribed by the delta function $\delta\left(\eta+\frac{\underline{\mathbf{x}}\cdot\underline{\mathbf{l}}}{c}\right)$. $\underline{\mathbf{x}}\equiv \underline{\mathbf{u}}\lambda$ is the baseline vector in units of distance. The details of the derivation of equation~(\ref{eqn:main:PS_FG_mm}) have been laid out in \S\ref{app:FG}. 

Equation~(\ref{eqn:main:PS_FG_mm}) enables evaluation of foreground contamination in $k$--space, which depends on classical source confusion $P_{lm}^\textrm{FG}(\underline{\mathbf{l}})$, the primary beam pattern of the tile $W_{lm}^\textrm{P}(\underline{\mathbf{l}})$, Fourier response of bandpass weights $W_\eta^\textrm{B}(\eta)$ and synthesized beam $S_{lm}(\underline{\mathbf{l}})$. It expresses the foreground contamination that is in $(u,v,\eta)$--coordinates (Fourier space) in terms of quantities in $(l,m)$--coordinates (real space). This expression aids in understanding the {\it mode--mixing} aspect of foreground contamination. The {\it delta} function indicates that unsubtracted sources at different locations and their sidelobes contribute predominantly to specific values of $\eta$ in delay space (and corresponding $k_\parallel$--modes) given by $\eta\simeq \underline{\mathbf{x}}\cdot\underline{\mathbf{l}}/c = \underline{\mathbf{u}}\cdot\underline{\mathbf{l}}/f$. This is consistent with the findings of \citet{ved12}. The equation naturally leads to a wedge--shaped region, whose boundary is set by the {\it horizon limit} \citep{par12} in $(l,m)$--space given by $\left|\:\underline{\mathbf{l}}\:\right|^2=l^2+m^2=1$. The equation in $k$--space for the {\it horizon limit} (wedge boundary) is:
\begin{equation}\label{eqn:horizon_line}
  k_\parallel = \frac{H_0\,E(z)\,D_\textrm{M}(z)}{c\,(1+z)}\,k_\perp.
\end{equation}

Using a {\it natural} weighted synthesized beam, $S_{lm}(\underline{\mathbf{l}})$, primary beam, $W_{lm}^\textrm{P}(\underline{\mathbf{l}})$, obtained by a phased $4\times 4$ array of identical radiators, instrumental delay response $W_\eta^\textrm{B}(\eta)$ for different bandpass shapes, and source confusion variance, $P_{lm}^\textrm{FG}(\underline{\mathbf{l}})$, we have estimated foreground contamination of power spectrum in full three--dimensional $k$--space by evaluating the integral in equation~(\ref{eqn:main:PS_FG_mm}). The size of a cell in three--dimensional $k$--space, termed {\it voxel}, corresponds to $\sim A_\textrm{e}$ and $\sim B_\textrm{eff}^{-1}$ in the transverse and line-of-sight directions respectively. The $k$--space volume of a {\it voxel} is $\simeq 9.1\times 10^{-8}$~Mpc$^{-3}$.

\begin{figure}[htb]
\centering
\includegraphics[width=\linewidth]{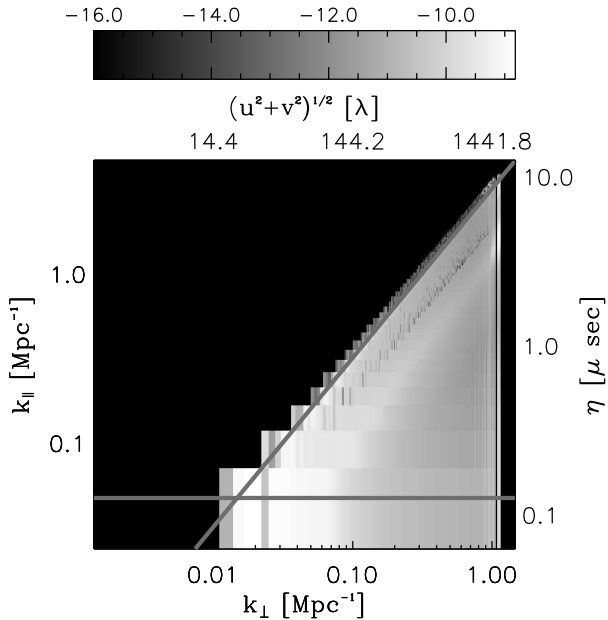}
\caption{Logarithm of azimuthally averaged foreground contamination (in units of K$^2$) in $k$--space for 6 hours of synthesis when bandpass effects are not applied. The gray horizontal line denotes the minimum $k_\parallel$ ($0.048$~Mpc$^{-1}\sim 1/B_\textrm{eff}$) mode that can be observed in the EoR H{\sc i} power spectrum. The upper limit on $k_\parallel$ is 4.7~Mpc$^{-1}$ ($\sim 1/2\,\Delta f$). The range in $k_\perp$ is 0.0014~Mpc$^{-1}\leq k_\perp\leq 1.47$~Mpc$^{-1}$, which is set by the range in baseline lengths ($\simeq 3$--$1636\lambda$). The gray line with positive slope defining the wedge boundary denotes the {\it horizon limit} within which the foreground power is predominantly contained in. The slope of the wedge is 3.18. The top axis denotes baseline lengths, $\left|\:\underline{\mathbf{u}}\:\right|\sim k_\perp$, in units of wavelength. The axis on the right denotes instrumental delay, $\eta\sim k_\parallel$. The grayscale color bar used is in logarithm units. Maximum contamination $\simeq 1.5\times 10^{-9}$~K$^2$ occurs below the instrumental window. Contamination at $(k_\perp,k_\parallel)\simeq (0.015,0.047)$~Mpc$^{-1}$ is typically $\sim 10^{-10}$~K$^2$. Contamination is enhanced close and parallel to the {\it horizon limit} because the confusion variance increases owing to rise in solid angles subtended. The black vertical segments near the right edge of the image indicate absence of measurements at the corresponding baselines. \label{fig:FG_PS_6_hrs_ideal}}
\end{figure}

In order to illustrate the three--dimensional foreground contamination in two--dimensional ($k_\perp,k_\parallel$)--plane, we azimuthally averaged the foreground contamination using an inverse quadrature weighting as though foregrounds were the only source of uncertainty. Figure~\ref{fig:FG_PS_6_hrs_ideal} shows the averaged foreground contamination (in units of $K^2$) in $k$--space for a 6~hour synthesis, where the effects of band shape, $W_\eta^\textrm{B}(\eta)$, are not applied. The structure of the foreground power spectrum is found to be in agreement with the foreground window illustrated in Figure~\ref{fig:kspace}. The lower and upper bounds on the $k_\perp$--axis are provided by the minimum ($\simeq 3\lambda$) and maximum ($\simeq 1636\lambda$) baseline lengths, and those on the $k_\parallel$--axis are provided by $\sim 1/B_\textrm{eff}$ (gray horizontal line, $\eta_\textrm{min}\simeq 0.125\,\mu$sec) and $\sim 1/(2\,\Delta f)$ ($\eta_\textrm{max}\simeq 12.5\,\mu$sec) respectively. The foreground component of the power spectrum occupies a wedge--shaped region in $k$--space. The {\it horizon limit} (gray line with positive slope) sets the boundary of the wedge. For the parameters listed in Table~\ref{tab:inst_parms}, the instrumental window is given by $0.0014$~Mpc$^{-1}\leq k_\perp\leq 1.47$~Mpc$^{-1}$ and $0.048$~Mpc$^{-1}\leq k_\parallel\leq 4.7$~Mpc$^{-1}$, and the slope of the foreground wedge is 3.18. The maximum foreground contamination is $\approx 1.5\times 10^{-9}$~K$^2$ but occurs below the instrumental window. The typical contamination close to $(k_\perp,k_\parallel)\simeq (0.015,0.047)$~Mpc$^{-1}$ is $\sim 10^{-10}$~K$^2$. The model EoR H{\sc i} signal strength in this region is $\sim 3\times 10^{-10}$~K$^2$. An apparent increase in foreground contamination closer and parallel to the wedge is noted. This is attributed to an increase in $\sigma_\textrm{C}^2(\underline{\mathbf{l}})$, which in turn is due to an increase in solid angles as $\left|\:\underline{\mathbf{l}}\:\right|$ approaches the {\it horizon limit}. 

If independent power spectra are averaged from $N_\textrm{fields}$ patches of sky, the foreground contamination in the averaged power spectrum goes as $\sim 1/\sqrt{N_\textrm{fields}}$. 

\subsection{Role of Bandpass Shapes in Foreground Contamination}\label{sec:BP_shaping}

An important consequence of convolution in equation~(\ref{eqn:main:PS_FG_mm}) by the term $W^\textrm{B}_\eta(\eta)$, the response of the bandpass shape in instrumental delay space, is to spill the contamination from foreground emission, which is restricted to the wedge, into the regions beyond. This spillover, therefore, fills even the desirable portions of $k$--space, namely, the {\it EoR window}. Hence, a simple removal of the wedge shaped region from data analysis does not completely remove all the effects of foreground contamination. The level of this spillover may be controlled by appropriate choice of bandpass shapes. \citet{ved12} have discussed a possibility of using a {\it Blackman--Nuttall} window function \citep{nut81}. 

In the context of bandpass window shaping, this paper addresses the following questions: what is the effect of bandpass shaping on foreground contamination in three--dimensional $k$--space? And, what is its significance to overall sensitivity when other uncertainties are also taken into account?

An infinite bandpass will convolve the power spectrum of foregrounds by a {\it delta} function resulting in zero spillover, but is impossible to achieve in practice. A more practical bandshape, such as a rectangular window, will manifest as a {\it sinc}--shaped response along $\eta$ (and $k_\parallel$). A {\it Blackman--Nuttall} window is known to have a much reduced sidelobe response (3--4 orders of magnitude) along $\eta$ but its peak (area under band shape) in $\eta$--space is $\approx 2.76$ times less than that of a {\it sinc} function, which implies a loss of sensitivity. The resolution in $\eta$ is also relatively poorer. However, if the effective bandwidth, $B_\textrm{eff}$, is made equal to that of a rectangular band shape by extending the standard {\it Blackman--Nuttall} window, a significant reduction in sidelobes in the response function along $\eta$ may be achieved without compromising either sensitivity or resolution, compared to those from a rectangular window. We define the effective bandwidth as:
\begin{equation}\label{eqn:eff_BW}
  B_\textrm{eff}=\int\limits_{-B_0/2}^{B_0/2}W_f^\textrm{B}(f)\,\dif f = W_\eta^\textrm{B}(0),
\end{equation}
where the second equality comes from the Fourier transform convention and the limits of the integral form the edges of the band.

\begin{figure}[htb]
\centering
\includegraphics[width=0.9\linewidth]{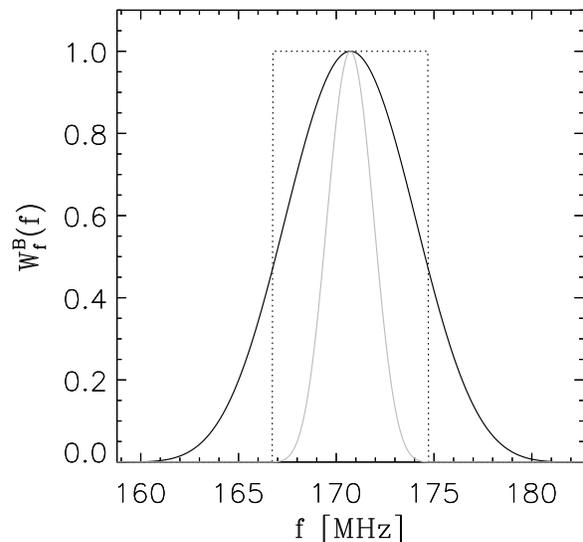}
\caption{Bandpass window functions centered at 170.7~MHz. The dotted curve is a rectangular window, the solid gray curve represents a standard {\it Blackman--Nuttall} window and the solid black curve represents an extended {\it Blackman--Nuttall} window whose effective bandwidth ($B_\textrm{eff}\sim$8~MHz) is equal to that of the rectangular window. For the standard {\it Blackman--Nuttall} window, $B_\textrm{eff}\sim$2.9~MHz. \label{fig:BP_windows}}
\end{figure}

\begin{figure}[htb]
\centering
\includegraphics[width=0.9\linewidth]{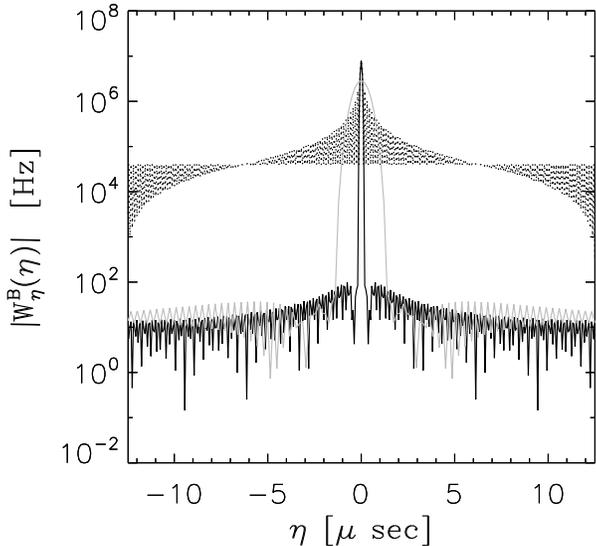}
\caption{The amplitude of the responses of the bandpass window functions of Figure~\ref{fig:BP_windows} shown versus $\eta$, the instrumental delay. Line styles are identical to those in Figure~\ref{fig:BP_windows}. The responses of rectangular and extended {\it Blackman--Nuttall} bandpass windows are almost identical in sensitivity and resolution along $\eta$, while the standard {\it Blackman--Nuttall} window is poorer in resolution and $\approx$~2.76 times lesser in sensitivity. 
  \label{fig:BP_response}}
\end{figure}

Figure~\ref{fig:BP_windows} shows the rectangular (dotted), standard (solid gray), and extended (solid black) versions of {\it Blackman--Nuttall} band shapes. Their effective bandwidths are 8~MHz, 2.9~MHz and 8~MHz respectively. Figure~\ref{fig:BP_response} shows using the respective line styles the amplitude of the respective responses, $\left|W_\eta^\textrm{B}(\eta)\right|$, of the aforementioned band shapes along $\eta$. As expected, the standard {\it Blackman--Nuttall} window has reduced sensitivity visible by its peak and is of a poorer resolution. The extended version, however, is identical in sensitivity and resolution to that of a rectangular window. 

Using $W_\eta^\textrm{B}(\eta)$ for rectangular and extended {\it Blackman--Nuttall} windows described above, the power spectrum of unsubtracted foreground sources was estimated in three--dimensional $k$--space using the integral in equation~(\ref{eqn:main:PS_FG_mm}). 

\begin{figure}[htb]
\centering
\includegraphics[width=\linewidth]{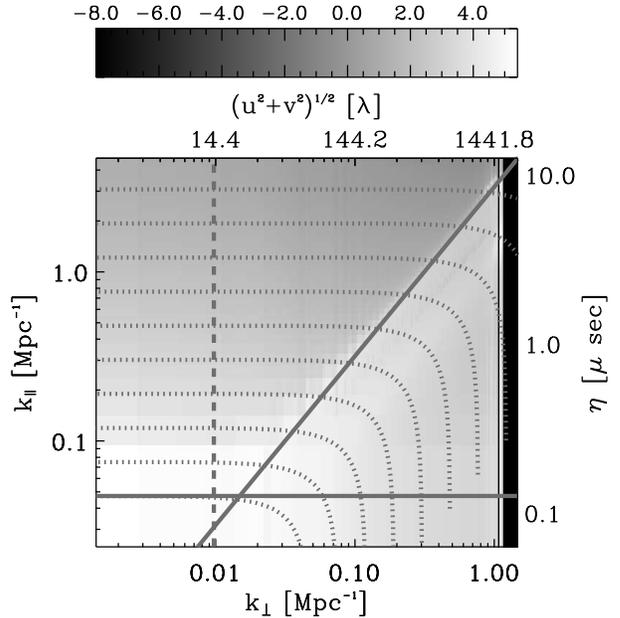}
\caption{Logarithm of azimuthally averaged foreground power spectrum (in units of K$^2$~Hz$^2$) with 6~hours of synthesis using a rectangular window. The solid gray lines are identical to those in Figure~\ref{fig:FG_PS_6_hrs_ideal}. The spillover of foreground power beyond the wedge is due to the response of rectangular bandpass window. The spillover into the {\it EoR window} around $(k_\perp,k_\parallel)\simeq (0.01,0.2)$~Mpc$^{-1}$ is $\sim 10^3$~K$^2$~Hz$^2$. The gray dotted lines denote boundaries of bins of $k=(k_\perp^2+k_\parallel^2)^{1/2}$, inside which the signal and uncertainties are averaged to obtain sensitivity. The grayscale color bar used is is logarithm units. A slice of the foreground power spectrum is obtained along the gray dashed line to estimate the spillover level beyond the wedge. The black vertical segments in the right edge of the image indicate absence of measurements at the corresponding baselines. \label{fig:FG_PS_6_hrs_RECT}}
\end{figure}

\begin{figure}[htb]
\centering
\includegraphics[width=\linewidth]{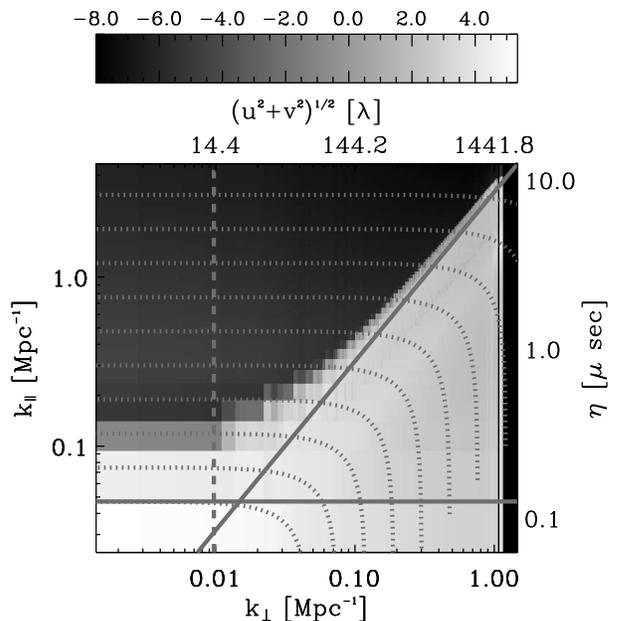}
\caption{Same as Figure~\ref{fig:FG_PS_6_hrs_RECT} but for an extended {\it Blackman--Nuttall} bandpass window. The spillover into the {\it EoR window} around $(k_\perp,k_\parallel)\simeq (0.01,0.2)$~Mpc$^{-1}$ is $\sim 10^{-5}$~K$^2$~Hz$^2$. \label{fig:FG_PS_6_hrs_BNW_EXT}}
\end{figure}

\begin{figure}[htb]
\centering
\includegraphics[width=0.9\linewidth]{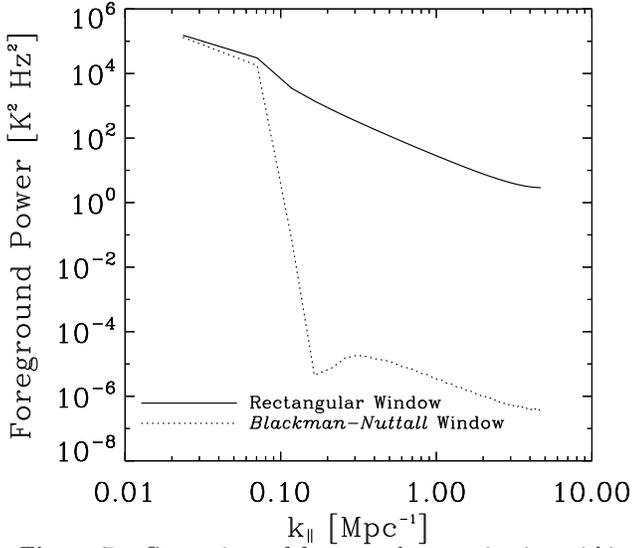}
\caption{Comparison of foreground contamination within the wedge and the spillover beyond for rectangular (solid line) and extended {\it Blackman--Nuttall} (dotted line) bandpass windows. The spillover is estimated along the slices shown as gray dashed lines in Figures~\ref{fig:FG_PS_6_hrs_RECT} and \ref{fig:FG_PS_6_hrs_BNW_EXT}. The spillover due to the latter is 7--8 orders of magnitude lower in the range 0.2~Mpc$^{-1}\lesssim k_\parallel\lesssim 5$~Mpc$^{-1}$ at $k_\perp\simeq 0.01$~Mpc$^{-1}$.\label{fig:slice_FG_PS_6_hrs}}
\end{figure}

Once again for purposes of illustration we averaged these power spectra in independent cells azimuthally in $k$--space using inverse quadrature weighting. Figures~\ref{fig:FG_PS_6_hrs_RECT} and \ref{fig:FG_PS_6_hrs_BNW_EXT} show the azimuthally averaged power spectra of foregrounds (in units of K$^2$~Hz$^2$) for rectangular and extended {\it Blackman--Nuttall} band shapes respectively, each with an effective bandwidth of 8~MHz. The features already noted in Figure~\ref{fig:FG_PS_6_hrs_ideal} are also noted here. But the bandpass effects were not applied in Figure~\ref{fig:FG_PS_6_hrs_ideal}. Due to the presence of the term $W_\eta^\textrm{B}(\eta)$ in equation~(\ref{eqn:main:PS_FG_mm}), the foreground contamination is seen to spill over into the {\it EoR window} in both cases. The spillover into the {\it EoR window} up to $k_\parallel < 0.1$~Mpc$^{-1}$ is similar in both cases and is $\sim 10^4$~K$^2$~Hz$^2$. A {\it Blackman--Nuttall} window is expected to be superior by 7--8 orders of magnitude in reducing the spillover beyond the wedge shaped region relative to a rectangular window because of the term $\left|W_\eta^\textrm{B}(\eta)\right|^2$. Higher levels of foreground contamination are seen due to a rectangular bandpass window in the region $k_\perp\lesssim 0.03$~Mpc$^{-1}, k_\parallel\gtrsim 0.1$~Mpc$^{-1}$ when compared to that due to a {\it Blackman--Nuttall} window. In fact, this is confirmed from Figure~\ref{fig:slice_FG_PS_6_hrs}, which compares the foreground power along slices at $k_\perp\simeq 0.01$~Mpc$^{-1}$ shown in Figures~\ref{fig:FG_PS_6_hrs_RECT} and \ref{fig:FG_PS_6_hrs_BNW_EXT} as gray dashed lines. In the range 0.2~Mpc$^{-1}\lesssim k_\parallel \lesssim 5$~Mpc$^{-1}$, the extended {\it Blackman--Nuttall} window produces a foreground contamination spillover $\sim 10^{-6}$--$10^{-5}$~K$^2$~Hz$^2$, which is about 7--8 orders of magnitude smaller than that from a rectangular window.

Are there undesirable effects of using an extended {\it Blackman--Nuttall} band shape? A wideband observation might be analyzed with a sliding window to examine for any change in EoR detection with redshift. Thus, bandwidth is not discarded when an extended {\it Blackman--Nuttall} window is deployed. But such a window uses larger total bandwidth (more channels) than a nominal rectangular window to achieve the same effective bandwidth. If there is significant cosmic evolution of the EoR signal within the band, the assumption of statistical stationarity of EoR signal could break down and lead to a dilution of measured signal power. 

\section{Thermal Noise Power Spectrum}\label{sec:TN}

Thermal noise component in a sampled visibility measurement in Fourier space, from equations~(\ref{eqn:obsvis}) and (\ref{eqn:obsvis_fourier}), is:
\begin{align}
V_{uv\eta}^\textrm{N}(\overline{\mathbf{u}}) &= \int V_{uvf}^\textrm{N}(\underline{\mathbf{u}},f)\,S_{uv}(\underline{\mathbf{u}})\,W_f^\textrm{B}(f)\,e^{-j2\pi\eta f}\dif f.
\end{align}

The rms of thermal noise in a measured visibility sample in a single frequency channel is given by \citep{mor05,mcq06}:
\begin{align}
  \Delta V_{uvf}^\textrm{N}(\underline{\mathbf{u}},f) &= \frac{\lambda^2\,T_\textrm{sys}}{A_\textrm{e}\sqrt{\Delta f\,t_\textrm{int}}}, \label{eqn:rms_vis_TN} 
\end{align}
where $t_\textrm{int}$ is the integration time used to obtain visibility samples.

In a {\it natural} weighting scheme, the weight of a certain ($u,v$)--cell is proportional to the number of baselines (or measurements) in that cell. When data measured by baselines inside ($uv$)--cells are averaged, the thermal noise in the averaged cell visibility is inversely proportional to the square root of number of baselines sampling that cell. Thus, the power spectrum uncertainty due to thermal noise in a spatial frequency mode ($\overline{\mathbf{u}}$) may be written as \citep{mor05,mcq06}:
\begin{align}
  C^\textrm{N}(\overline{\mathbf{u}}) &= \left(\frac{\lambda^2\,T_\textrm{sys}}{A_\textrm{e}}\right)^2\frac{\epsilon B_\textrm{eff}}{t_{\underline{\mathbf{u}}}}, \label{eqn:TN_PS}
\end{align}
where $\epsilon$ is the factor arising out of bandpass weights and $t_{\underline{\mathbf{u}}}$ is the effective time of observation for a given mode, which is effectively a product of the number of baselines sampling the mode during aperture synthesis and the integration time used in producing a visibility sample. 

Thermal noise power spectrum is estimated as prescribed by \citet{mor05} and \citet{mcq06}. $\epsilon$ arises in equation~(\ref{eqn:TN_PS}) from the sum of squares of bandpass weights, $W_f^\textrm{B}(f)$, while determining the power spectrum. $\epsilon$ is given by:
\begin{align}
  \epsilon &= \frac{\displaystyle\sum_{i=1}^{N_\textrm{ch}}\,\big|W_f^\textrm{B}(f_i)\big|^2}{\displaystyle\sum_{i=1}^{N_\textrm{ch}}\,W_f^\textrm{B}(f_i)}, 
\end{align}
where $i$ indexes the frequency channels in the bandpass, $N_\textrm{ch}$ is the number of channels in the bandpass and $\sum_{i=1}^{N_\textrm{ch}}\,W_f^\textrm{B}(f_i)\,\Delta f = B_\textrm{eff}$. The value of $\epsilon$ is 1 and 0.72 respectively for rectangular and extended {\it Blackman--Nuttall} window band shapes. It indicates that an extended {\it Blackman--Nuttall} window also reduces the thermal noise component in the power spectrum by 28\%. As a result, in order to achieve a thermal noise power equal to that from an extended {\it Blackman--Nuttall} window using a rectangular window, an observation has to be $\approx 40$\% longer.

Uncertainty in power spectrum due to thermal noise is very sensitive to observing time. In each $k$--mode, it is inversely proportional to the number of baselines, including redundant ones that sample that $k$--mode during the entire synthesis. It depends on band shape that is parametrized by $\epsilon$. In the context of our model observations listed in Table~\ref{tab:obs_parms}, the thermal noise power spectrum in equation~(\ref{eqn:TN_PS}) at any $\overline{\mathbf{k}}$--mode may be re-written as:
\begin{align}\label{eqn:TN_PS_obs_2D}
  C^\textrm{N}(\overline{\mathbf{k}}) = \left(\frac{\lambda^2\,T_\textrm{sys}}{A_\textrm{e}}\right)^2\frac{\epsilon\, B_\textrm{eff}}{N_\textrm{cad}\,t_\textrm{int}\,N(\underline{\mathbf{k}})\,\sqrt{N_\textrm{fields}}},
\end{align}
where $N(\underline{\mathbf{k}})$ is the number of baselines (redundant ones included) observing the $\underline{\mathbf{k}}$--mode during a single synthesis observation of duration $t_\textrm{syn}$ and $N_\textrm{cad}=t_\textrm{obs}/(t_\textrm{syn}\,N_\textrm{fields})$. 

\begin{figure}[htb]
\centering
\includegraphics[width=\linewidth]{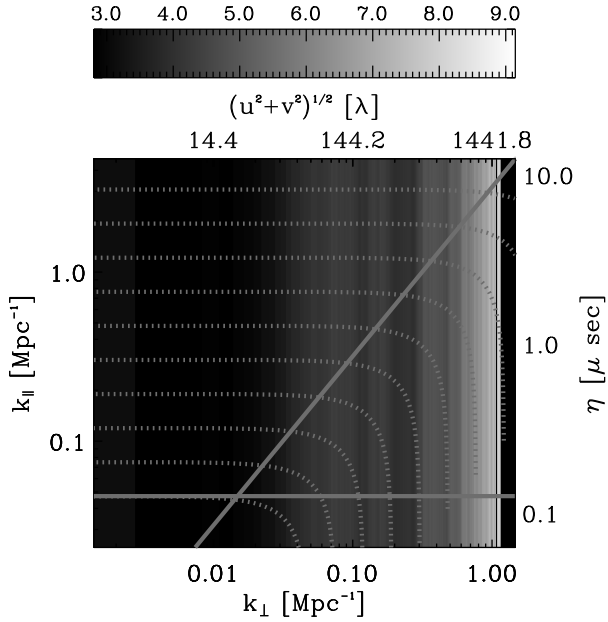}
\caption{Logarithm of thermal power spectrum (in units of K$^2$~Hz$^2$) for a 1000~hour observation with 6~hours synthesis on a single field (first observing mode in Table~\ref{tab:obs_parms}), using a rectangular bandpass window ($\epsilon=1$). The solid and dotted gray lines are identical to those shown in Figure~\ref{fig:FG_PS_6_hrs_RECT}. The grayscale color bar used is in logarithm units. The minimum ($\approx 640$~K$^2$~Hz$^2$) and maximum ($\approx 1.4\times 10^9$~K$^2$~Hz$^2$) are attained at $k_\perp\simeq 0.004$~Mpc$^{-1}$ and $k_\perp\simeq 1$~Mpc$^{-1}$ respectively for the {\it natural} weighting scheme used. The black vertical segments in the right edge of the image indicate absence of measurements at the corresponding baselines. \label{fig:TN_PS_2D_1000h_single_RECT}}
\end{figure}

\begin{figure}[htb]
\centering
\includegraphics[width=\linewidth]{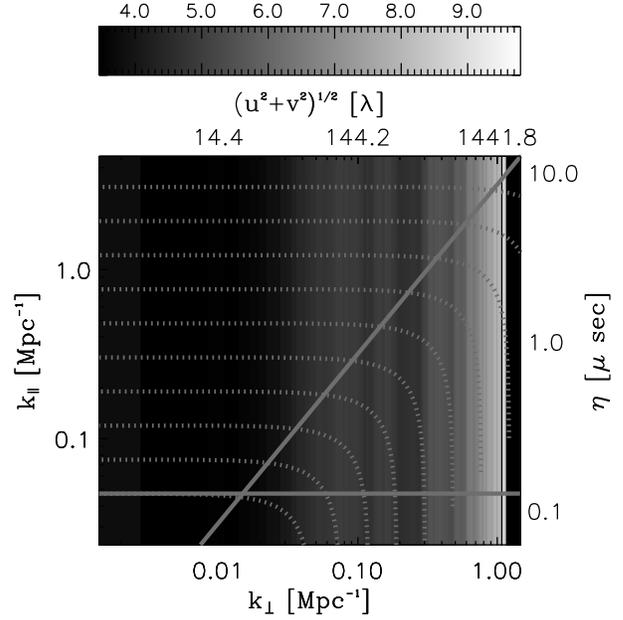}
\caption{Same as Figure~\ref{fig:TN_PS_2D_1000h_single_RECT} but evaluated for 1000~hours with 6~hours synthesis on 20 fields (observing mode~(2) in Table~\ref{tab:obs_parms}). In this case, thermal noise power spectrum relative to observing mode~(1) (in Figure~\ref{fig:TN_PS_2D_1000h_single_RECT}) is worse by a factor $\approx 4.47$ throughout, owing to a decrease in observing time on individual fields as the number of independent fields observed is increased from one to twenty. \label{fig:TN_PS_2D_1000h_multi_RECT}}
\end{figure}

For illustration, this is azimuthally averaged in bins of $k_\perp=(k_x^2+k_y^2)^{1/2}$ by adding the thermal noise component of power spectrum contained in cells in this bin, in inverse quadrature, to yield a two--dimensional thermal noise power spectrum, $C^\textrm{N}(k_\perp,k_\parallel)$. In Figures~\ref{fig:TN_PS_2D_1000h_single_RECT} and \ref{fig:TN_PS_2D_1000h_multi_RECT}, we show $C^\textrm{N}(k_\perp,k_\parallel)$ for observing modes~(1) and (2), respectively, using a rectangular bandpass window. In observing mode~(1), the azimuthally averaged thermal noise power spectrum attains minimum ($\approx 640$~K$^2$~Hz$^2$) at $k_\perp\simeq 0.004$~Mpc$^{-1}$ and maximum ($\approx 1.4\times 10^9$~K$^2$~Hz$^2$) at $k_\perp\simeq 1$~Mpc$^{-1}$. The thermal noise power spectra are similar in the two observing modes except that it is higher by a factor of $\approx 4.47$ in the latter as predicted by the scaling in equation~(\ref{eqn:TN_PS_obs_2D}). In the first observing mode, the visibilities are added coherently for a total of 1000~hours. In the second observing mode, the visibilities on each field are added coherently only for 50~hours and power spectra are estimated. These power spectra are then averaged. This increases the thermal noise component in the power spectrum by a factor $\sqrt{20}\approx 4.47$.

Thermal noise power spectrum exhibits cylindrical symmetry about the $k_\parallel$--axis. When an extended {\it Blackman--Nuttall} window is used, thermal noise component in the power spectrum drops by 28\% ($\epsilon=0.72$) relative to that from a rectangular bandpass window.

\section{EoR H{\sc i} Power Spectrum and Sample Variance}\label{sec:EoR_PS_SV}

The measured H{\sc i} component of Fourier space visibilities in equation~(\ref{eqn:obsvis}) in a manner similar to equation~(\ref{eqn:main:FG_vis_gen}) may be expressed as:
\begin{align}\label{eqn:main:HI_vis_gen}
  V_{uv\eta}^\textrm{H{\sc i}}(\overline{\mathbf{u}}) &= V_{uv\eta}^\textrm{H{\sc i};T}(\overline{\mathbf{u}})\,S_{uv}(\underline{\mathbf{u}}) \ast W_{uv\eta}(\overline{\mathbf{u}}).
\end{align}
The EoR H{\sc i} power spectrum measured by the instrument, $P^\textrm{H{\sc i}}_\textrm{inst}(\overline{\mathbf{k}})$, is the diagonal of the covariance matrix $\Bigl\langle V_{uv\eta}^\textrm{H{\sc i}}(\overline{\mathbf{u}}_i)^\star V_{uv\eta}^\textrm{H{\sc i}}(\overline{\mathbf{u}}_j)\Bigr\rangle$ and is given by \citep{mor04,mor05,mcq06,bow06,bow07}:
\begin{align}\label{eqn:PS_HI_gen}
  P^\textrm{H{\sc i}}_\textrm{inst}(\overline{\mathbf{k}}) &= \iiint P^\textrm{H{\sc i}}(\overline{\mathbf{u^\prime}})\,\left|W_{uv\eta}(\overline{\mathbf{u}}-\overline{\mathbf{u^\prime}})\right|^2\,\dif^3 \overline{\mathbf{u^\prime}}
\end{align}
at the sampled baseline locations and $P^\textrm{H{\sc i}}(\overline{\mathbf{u}})$ is given by equation~(\ref{eqn:jacobian}). Although some authors \citep{mcq06,bow06,bow07} have approximated $W_{uv\eta}(\overline{\mathbf{u}})$ by a delta function, it will exhibit some spillover along $\eta$ depending on the bandpass shape $W_f^\textrm{B}(f)$. Hence, we retain the general form of the power spectrum in equation~(\ref{eqn:PS_HI_gen}) for our work.

In equation~(\ref{eqn:PS_HI_gen}), we note the convolving effect arising out of instrumental factors, thereby introducing correlations between neighboring spatial frequencies. The observed power spectrum of the signal is a modification of the true EoR H{\sc i} power spectrum by instrumental parameters of observation such as primary beam and bandpass shape. 

Sample variance is equal to the power spectrum \citep{jun96,mcq06}. If a number of independent measurements ($N_\textrm{fields}$) of power spectrum are averaged, the sample variance goes as $P^\textrm{SV}(\overline{\mathbf{k}})=P^\textrm{H{\sc i}}_\textrm{inst}(\overline{\mathbf{k}})/\sqrt{N_\textrm{fields}}$. 

$P^\textrm{H{\sc i}}(k)$, in equation~(\ref{eqn:EoR_PS}), represents the variance in $k$--space of the spin temperature fluctuations of H{\sc i} relative to the CMB. As already mentioned in \S\ref{sec:instr_parms}, simulations of \citet{lid08} show that the variance in the ionization field peaks at a value close to 50\% ionization. We choose from the family of $P^\textrm{H{\sc i}}(k)$ curves they provide the one parametrized by $(\langle x_\textrm{i}\rangle, z)=(0.54, 7.32)$ and use it as the input model in this study. $\langle x_\textrm{i}\rangle$ and $z$ are the mean volume-averaged ionization fraction and redshift respectively. Redshifted emission from H{\sc i} at $z=7.32$ occurs at 170.7~MHz, which is chosen as our observing frequency. We have obtained the values of power spectrum predicted for $(\langle x_\textrm{i}\rangle, z)=(0.54, 7.32)$ from the plots of \citet{lid08} (through private communication with Adam Lidz). Since we required the predicted values of power spectrum at intermediate values of $k$ not tabulated, we used a third--order polynomial fit to interpolate the predicted power spectrum to the required values.

Two primary causes contribute to power spectrum of EoR H{\sc i} fluctuations:
\begin{enumerate}
\item the underlying matter density fluctuations, and
\item the ionized bubbles during the reionization process. 
\end{enumerate}
The contribution from matter density fluctuations is anisotropic due to redshift--space distortions caused by peculiar velocity effects along the line of sight \citep{bar05}, whereas, the contribution from ionization fluctuations is isotropic. In our adopted model the ionization fraction is about 50\%, which indicates significant ionization. Hence, the contribution to the EoR can be assumed to be dominated by ionized bubbles rather than due to underlying matter density fluctuations \citep{lid08}. Therefore, we neglect anisotropic effects arising out of peculiar velocities in our model power spectrum. 

The observed power spectrum is computed from equation~(\ref{eqn:PS_HI_gen}) using the input model. It is identical for the two observing modes listed in Table~\ref{tab:obs_parms}. The observed sample variance is higher in observing mode~(1) relative to observing mode~(2) by a factor $\sqrt{20}\approx 4.47$. 

\begin{figure}[htb]
\centering
\includegraphics[width=\linewidth]{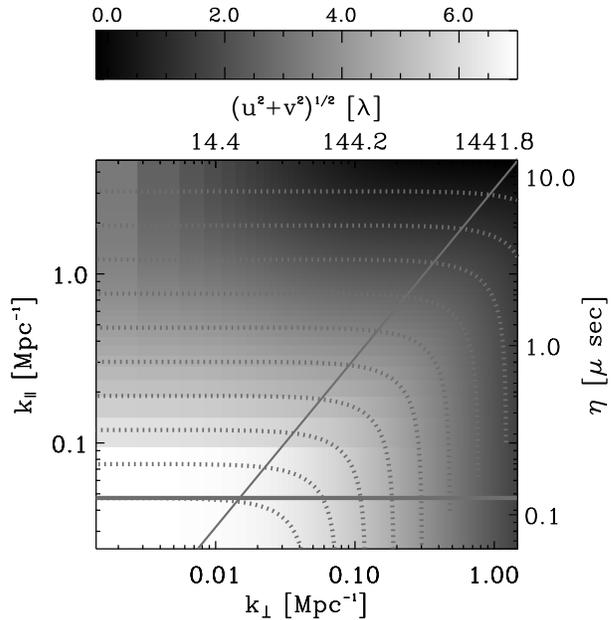}
\caption{Logarithm of EoR H{\sc i} signal power spectrum (in units of K$^2$~Hz$^2$) for observing mode~(1) listed in Table~\ref{tab:obs_parms} using a rectangular bandpass window. The solid and dotted gray lines are identical to those shown in Figure~\ref{fig:FG_PS_6_hrs_RECT}. The grayscale color bar used is in logarithm units.\label{fig:EoR_PS_obs_6_hrs_RECT}}
\end{figure}

Figure~\ref{fig:EoR_PS_obs_6_hrs_RECT} shows the observed EoR H{\sc i} power spectrum in ($k_\perp,k_\parallel$)--plane corresponding to observing mode~(1) while employing a rectangular band shape. Bandpass shape causes a convolution with the true power spectrum along $k_\parallel$, as in the case of foreground power spectrum. Although not shown, as expected the extended {\it Blackman--Nuttall} window causes a far lesser spillover of the EoR H{\sc i} power spectrum relative to the rectangular band shape. For our paper, we term this as ``signal spillover''. It is caused by the same reason (bandpass window shape) that causes foreground spillover beyond the {\it horizon limit}.  

\section{EoR Signal Detection}\label{sec:EoR_detection}

We have estimated the EoR H{\sc i} power spectrum expected to be observed and individual uncertainties in three--dimensional $k$--space. The total uncertainty in the power spectrum in three--dimensional $k$--space was obtained by summing the component uncertainties, 
\begin{equation}
  \Delta P(\overline{\mathbf{k}}) = P^\textrm{FG}_\textrm{inst}(\overline{\mathbf{k}}) + C^\textrm{N}(\overline{\mathbf{k}}) + P^\textrm{SV}(\overline{\mathbf{k}}).
\end{equation}

\subsection{Family of {\it EoR Windows}}\label{sec:EoR_windows}

By knowing the occupancy of various uncertainties in $k$--space and excluding these regions where uncertainties dominate, the estimates are expected to be relatively free of contamination \citep{mor12}. Within the instrumental window, results from \citet{dat10}, \citet{ved12}, \citet{wil12} and our study have shown that the wedge-shaped region in the $k$--space is contaminated due to unsubtracted foreground sources and their sidelobes relatively more than in other regions in $k$--space. Hence, the {\it EoR window} for H{\sc i} power spectrum has been designated as the region in $k$--space inside the instrumental window excluding the wedge. The idea of an optimal window is investigated further. 

We have shown that foregrounds are not strictly contained within the wedge-shaped region (see \S\ref{sec:BP_shaping}). A spillover from the wedge-shaped region is caused by the instrumental delay function $W_\eta^\textrm{B}(\eta)$. The characteristic width of this convolving instrumental response is proportional to $B_\textrm{eff}^{-1}$. Thus, immediately following the wedge boundary determined by the {\it horizon limit} given by equation~(\ref{eqn:horizon_line}), the spillover up to a few characteristic widths of convolution by the instrumental frequency response is also found to contain higher levels of contamination (see Figures~\ref{fig:FG_PS_6_hrs_ideal}, \ref{fig:FG_PS_6_hrs_RECT} and \ref{fig:FG_PS_6_hrs_BNW_EXT}). 

\begin{figure}[htb]
\centering
\includegraphics[width=\linewidth]{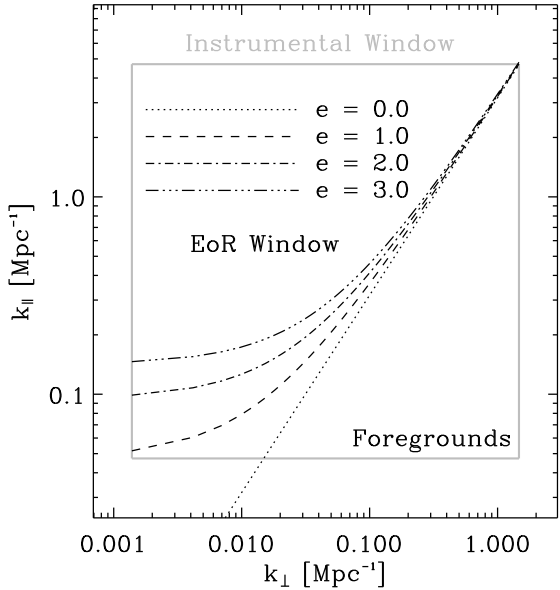}
\caption{Family of {\it EoR windows} parametrized by $e$ in equation~(\ref{eqn:EoR_windows}). The gray box denotes the instrumental window in $k$--space. The straight dotted line marks the boundary of the wedge-shaped region given by equation~(\ref{eqn:horizon_line}). $e$ denotes the number of characteristic widths ($\propto\, B_\textrm{eff}^{-1}$) of convolution due to the instrumental delay function $W_\eta^\textrm{B}(\eta)$. $e=0$ denotes only the wedge term is removed, while $e=3$ denotes removal of three characteristic widths in addition to the wedge. \label{fig:EoR_windows}}
\end{figure}

We investigate refinements to the so called {\it EoR window}. We narrow the {\it EoR window} by adding a term proportional to the characteristic convolution width. We define a refined {\it EoR window} as the region:
\begin{equation}\label{eqn:EoR_windows}
  k_\parallel \geq \frac{H_0\,E(z)\,D_\textrm{M}(z)}{c\,(1+z)}\left(k_\perp + \frac{e}{B_\textrm{eff}}\,\frac{2\pi\,f_{21}}{(1+z)\,D_\textrm{M}(z)}\right).
\end{equation}
This will reduce to the {\it horizon limit} in equation~(\ref{eqn:horizon_line}) without the second term in parenthesis. The second term is proportional to the characteristic width of the convolution arising from instrumental delay function $W_\eta^\textrm{B}(\eta)$. $e$ parametrizes this constant of proportionality. Equation~(\ref{eqn:EoR_windows}) represents a family of {\it EoR windows}. This concept is illustrated in Figure~\ref{fig:EoR_windows}. The instrumental window in $k$--space is shown as a gray box. The bottom right corner of this box is the region contaminated by foregrounds and the top left corner represents the refined {\it EoR window}.

\subsection{One--dimensional Sensitivity}\label{sec:sensitivity_1D}

Using the model for the EoR H{\sc i} power spectrum and with estimates of three primary uncertainties, namely, the foregrounds, thermal noise and sample variance, the sensitivity of the instrument to EoR power spectrum detection may be obtained. By averaging signal and uncertainties in independent {\it voxels} in spherical shells of $k=(k_\perp^2+k_\parallel^2)^{1/2}$, sensitivity may be improved. 

The model EoR H{\sc i} power spectrum we have considered is spherically symmetric and hence a function only of radial coordinate in $k$--space. This symmetry is modified to an extent by the instrument observing the power spectrum, because the instrumental term, $W_{uv\eta}(\overline{\mathbf{u}})$, in equation~(\ref{eqn:PS_HI_gen}) is not spherically symmetric. During spherical averaging, we ignore this loss of spherical symmetry caused by instrumental distortion. While averaging in shells of $k$, we average the observed signal in these shells and, correspondingly, estimate the uncertainties by adding them in inverse quadrature. 

We compare the one--dimensional signal and noise estimates for the cases listed in Table~\ref{tab:obs_parms}, while deploying rectangular and extended {\it Blackman--Nuttall} band shapes, for a range of values of $e$. 

\begin{figure*}[htb]
\centering
\subfloat[][Sensitivity for $e=0$]{\label{fig:SNR1D_6hrs_167_single_e0}\includegraphics[width=0.45\linewidth]{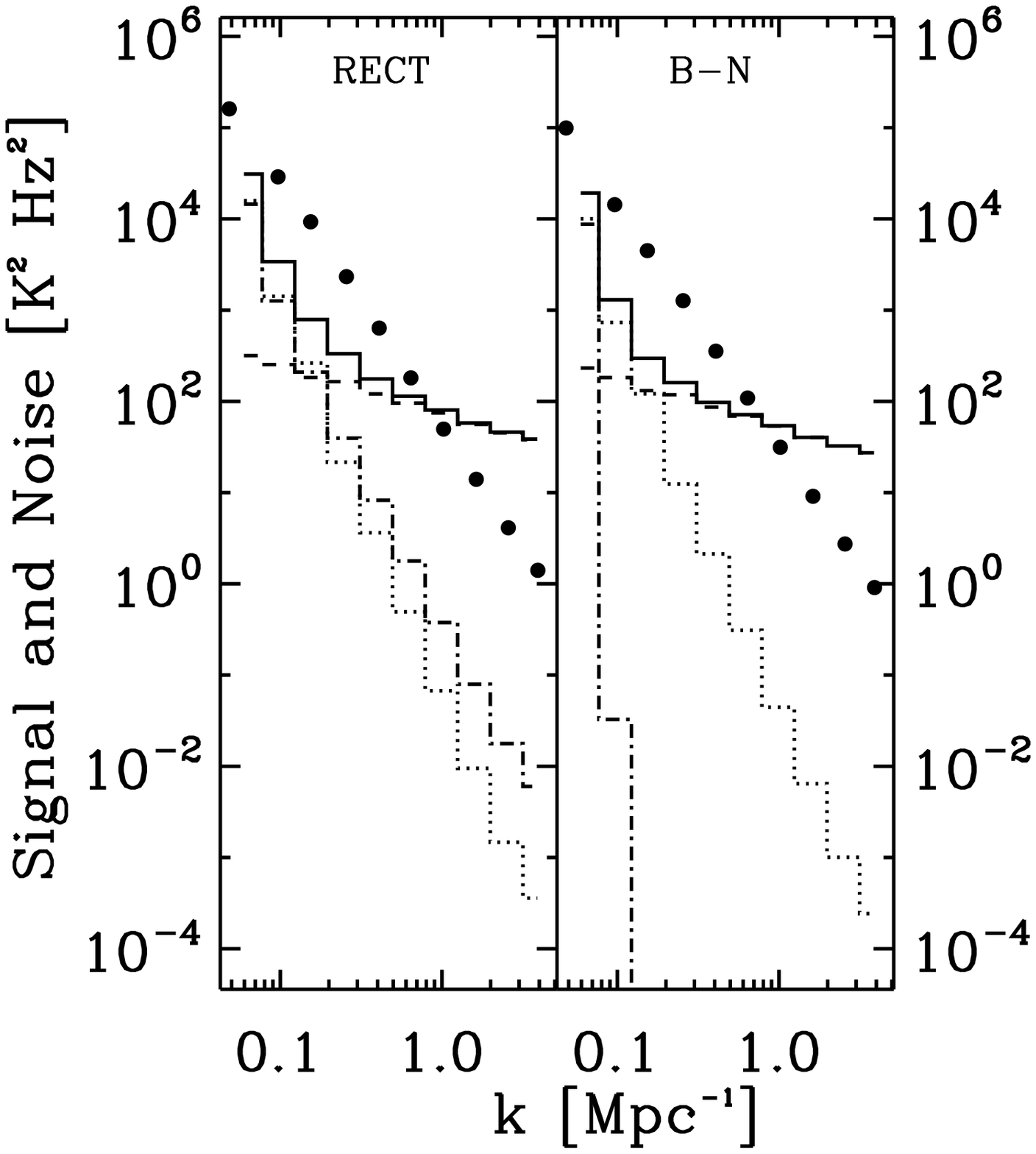}}
\subfloat[][EoR H{\sc i} and foregrounds]{\label{fig:case3:norm_EoR_FG_PS}\includegraphics[width=0.45\linewidth]{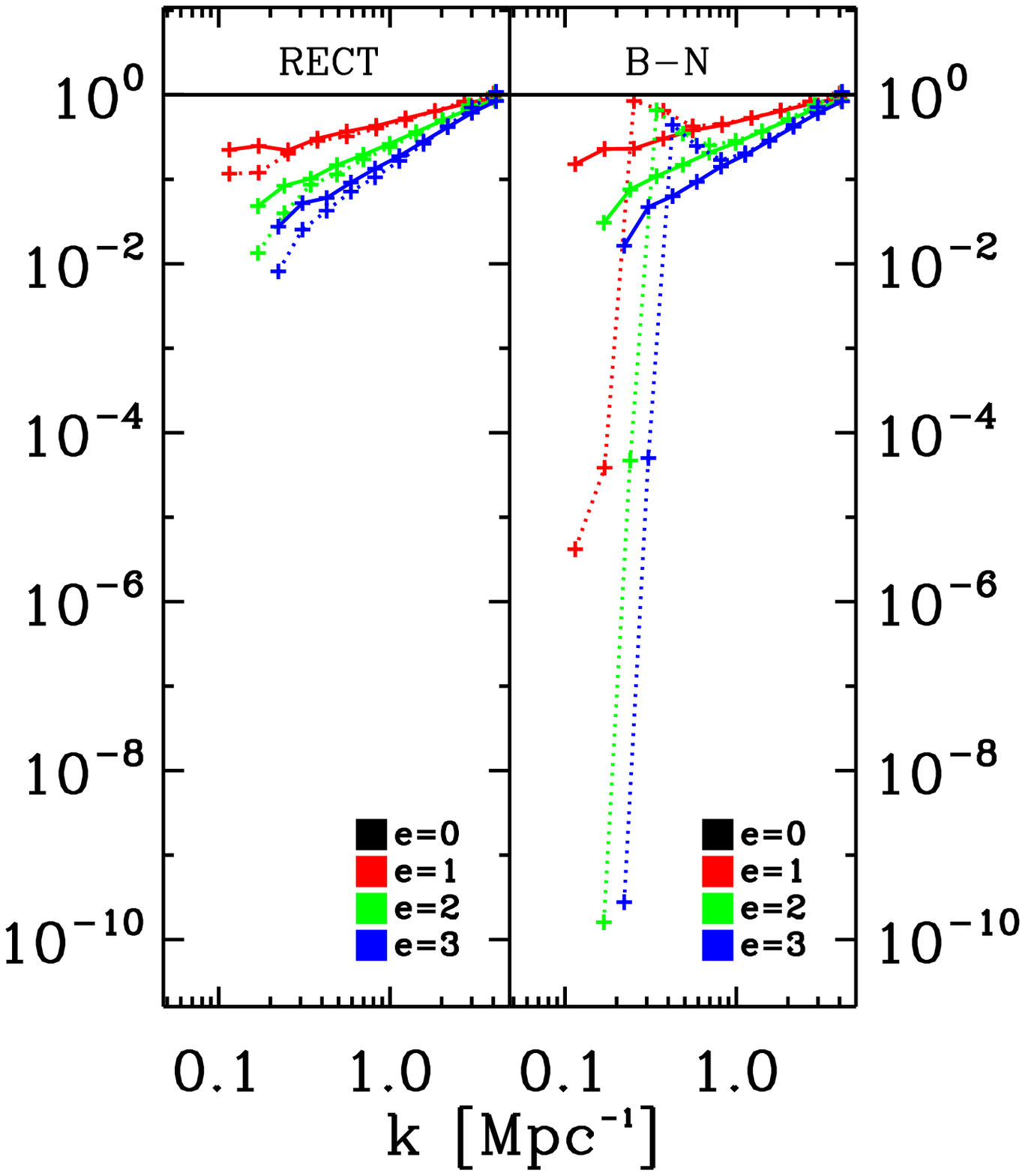}} \\
\subfloat[][Thermal noise]{\label{fig:case3:norm_TN_PS}\includegraphics[width=0.45\linewidth]{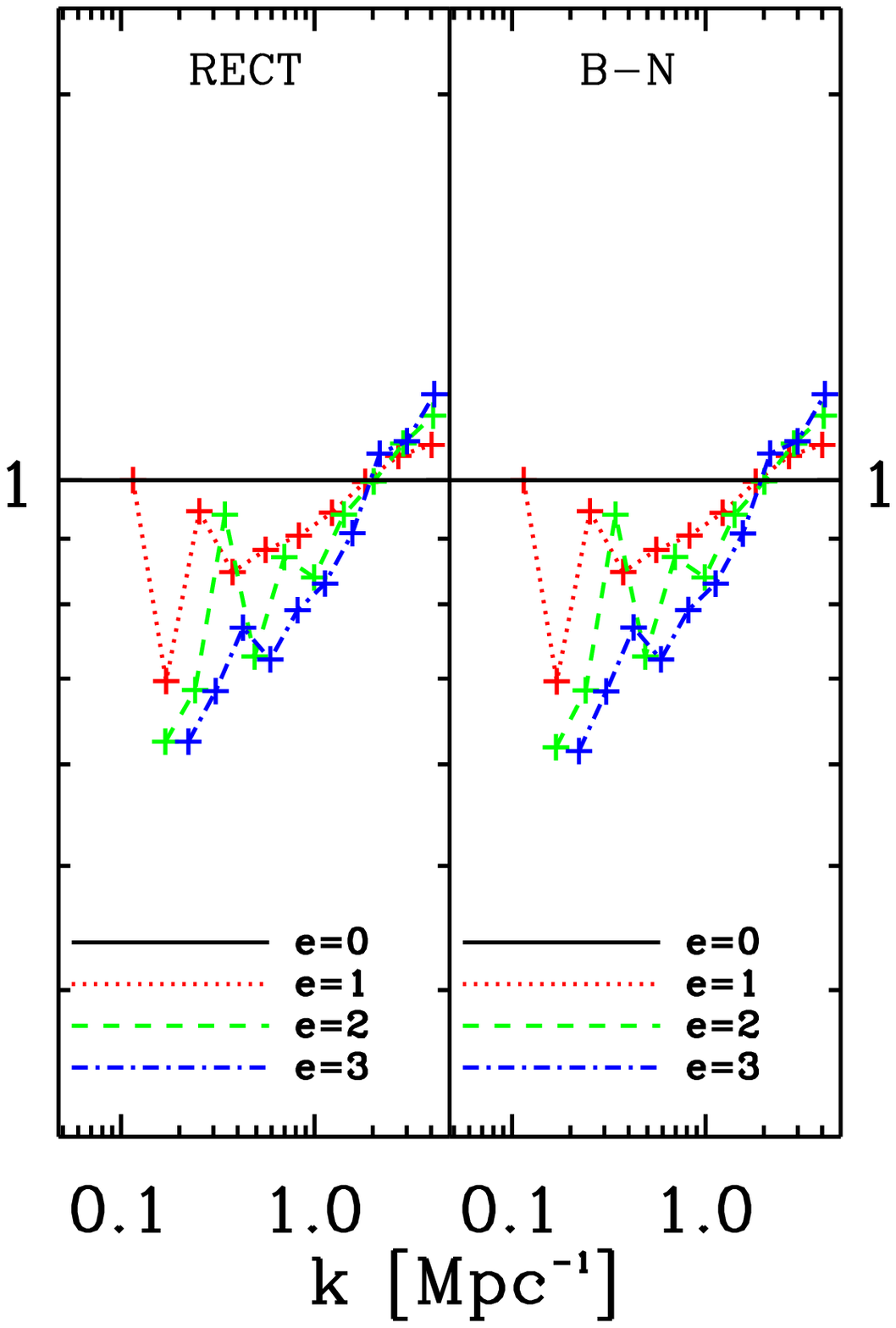}}
\subfloat[][S/N ratio]{\label{fig:case3:SNR}\includegraphics[width=0.45\linewidth]{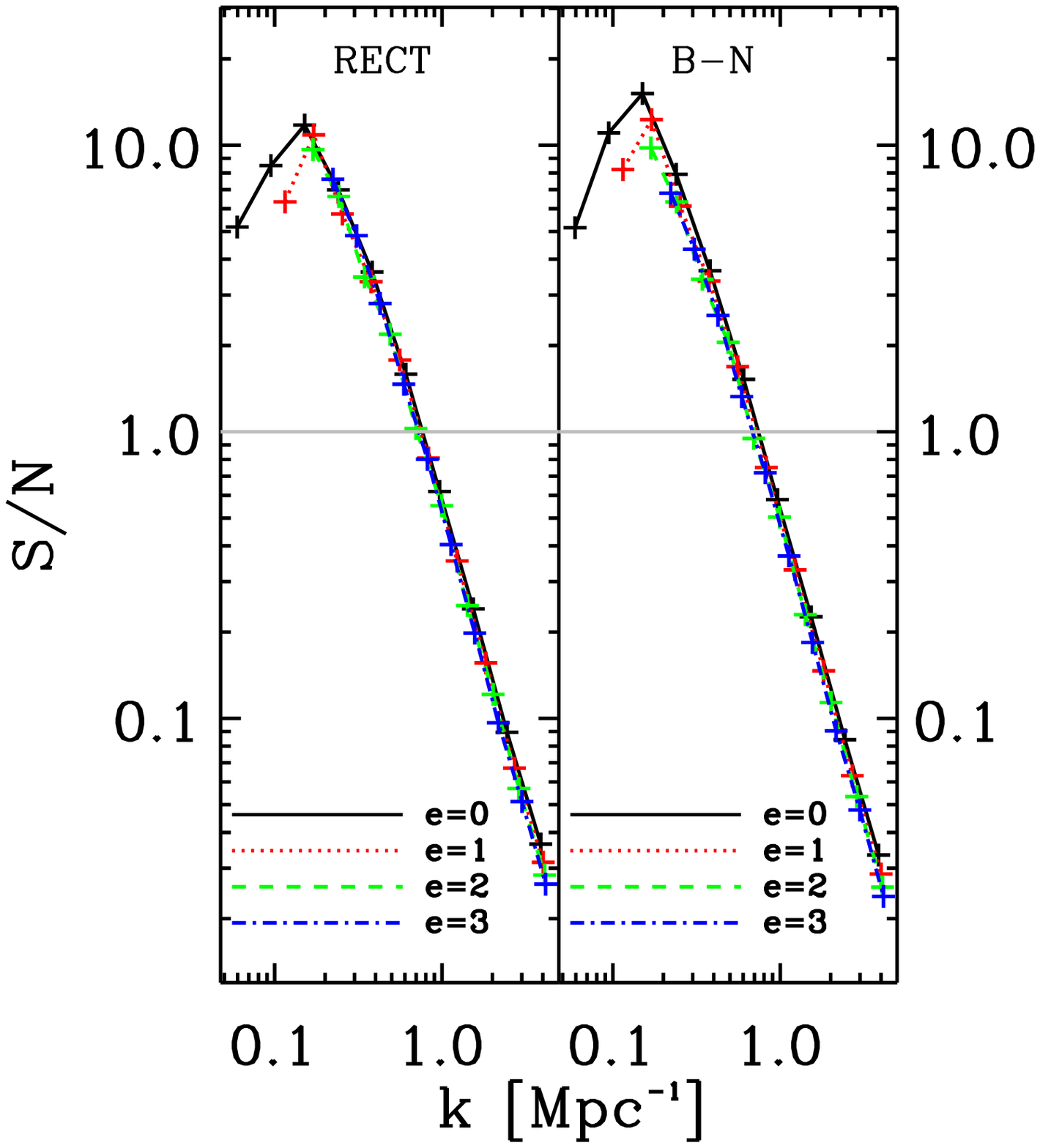}}
\caption{Properties of signal and uncertainty components in power spectrum (in units of K$^2$~Hz$^2$) for observing mode~(1). These are determined by averaging in shells of $k$ for various values of {\it EoR window} parameter, $e$. In each panel, the sub-panel on the left is obtained with a rectangular bandpass shape while that on the right is due to an extended {\it Blackman--Nuttall} window. The averaging excludes from the ($k_\perp,k_\parallel$)--plane, the wedge-shaped foreground window and an additional region parametrized by $e$ in equation~(\ref{eqn:EoR_windows}). ({\it a})~Expected EoR H{\sc i} signal and uncertainties in the power spectrum. The solid circles and lines denote the signal and the total uncertainty respectively. The latter consists of sample variance (dotted), thermal noise (dashed), and foreground contamination (dot dashed). 
  ({\it b})~Signal (solid lines) and foreground contamination (dotted lines) for different values of $e$ ($e=1,2,3$ in red, green and blue respectively) normalized by their respective values at $e=0$ (black) in Figure~\ref{fig:SNR1D_6hrs_167_single_e0}. ({\it c})~Thermal noise component in power spectrum for different values of $e$ (shown in legend) normalized by its values at $e=0$. ({\it d})~S/N ratios for different values of $e$. In panels {\it b} through {\it d}, `+' symbols denote the mean radii of spherical shells.}
\label{fig:SN_EoR_windows_case_3}
\end{figure*}

Figures~\ref{fig:SNR1D_6hrs_167_single_e0}--\ref{fig:case3:SNR} show in detail the signal and uncertainties expected with the MWA for observing mode~(1) listed in Table~\ref{tab:obs_parms}. Figure~\ref{fig:SNR1D_6hrs_167_single_e0} demonstrates the levels of signal (solid circles) and uncertainty (solid line) expected with either of the bandpass windows employed for the {\it EoR window} parameter $e=0$. Also shown are the individual components of the total uncertainty in different line styles (foregrounds: dot-dashed, thermal noise: dashed, sample variance: dotted). Figure~\ref{fig:case3:norm_EoR_FG_PS} shows the change in the power spectrum of EoR H{\sc i} and that of foregrounds when $e$ is varied ($e=1,2,3$ in red, green and blue respectively) relative to their respective values at $e=0$ (black). Similarly, Figure~\ref{fig:case3:norm_TN_PS} shows the change in thermal noise component of power spectrum for different values of $e$, relative to its values at $e=0$. In other words, Figures~\ref{fig:case3:norm_EoR_FG_PS} and \ref{fig:case3:norm_TN_PS} show signal and uncertainty components normalized with respect to themselves obtained at $e=0$. Hence, the quantities at $e=0$ are shown for reference in these Figures at a constant value of unity. Figure~\ref{fig:case3:SNR} shows the ratio of signal to uncertainty (S/N) as $e$ is varied. The left sub-panel in each panel corresponds to a rectangular bandpass window while that on the right is obtained with an extended {\it Blackman--Nuttall} window. 

With observing mode~(1) using 128--tile MWA, the signal clearly appears to be detectable (S/N~$>1$) for $k\lesssim 0.8$~Mpc$^{-1}$ (see Figures~\ref{fig:SNR1D_6hrs_167_single_e0} and \ref{fig:case3:SNR}). From Figure~\ref{fig:SNR1D_6hrs_167_single_e0}, with $e=0$, foreground contamination (dot-dashed line) exceeds thermal noise (dashed line) for $k\lesssim 0.2$~Mpc$^{-1}$ and $k\lesssim 0.1$~Mpc$^{-1}$ while using rectangular and {\it Blackman--Nuttall} windows respectively. Beyond this crossover, thermal noise takes over as the dominant source of uncertainty in power spectrum. Foregrounds (dot-dashed line) and sample variance (dotted line) are roughly equal up to this crossover. As $e$ increases, progressively larger regions get excluded from $k$--space. This is clearly visible in Figures~\ref{fig:case3:norm_EoR_FG_PS}--\ref{fig:case3:SNR}, especially for $k\lesssim 1$~Mpc$^{-1}$, through a systematic drift of radii of spherical shells (`+' symbols) towards higher values as $e$ increases (red to green to blue). In other words, increasing $e$ from 0 to 3 makes it progressively harder to recover scales with 0.06~Mpc$^{-1}\lesssim k\lesssim 0.2$~Mpc$^{-1}$. This results in partial removal of different uncertainties and the signal, besides an inherent decrease in signal strength with increasing $k$. However, the decrease in signal by $\lesssim$~1--2 orders of magnitude (solid red, green and blue curves) is less rapid than that in the foreground contamination (dashed red, green and blue curves), which decreases by $\gtrsim$~1--2 orders of magnitude. This is true for both bandpass shapes but is quite pronounced for the extended {\it Blackman--Nuttall} window (see Figure~\ref{fig:case3:norm_EoR_FG_PS}), where the foreground contamination reduces by $\sim$~5--10 orders of magnitude. On the other hand, as $e$ increases, thermal noise component changes  at most by a factor of 2 as seen from Figure~\ref{fig:case3:norm_TN_PS}. Effectively, the thermal noise component is only mildly affected compared to the signal and foregrounds. Regardless of the bandpass window used, the colored curves which are almost coincident in Figure~\ref{fig:case3:SNR} show that there is no improvement in overall sensitivity as $e$ is varied. This is a consequence of the nature of inverse quadrature weighting used in averaging in spherical shells of $k$. However, it is very important to reiterate that the foreground contamination decreases more rapidly than the loss in signal as $e$ is increased. In fact, foregrounds are almost completely removed for $e\gtrsim 1$ while using the {\it Blackman--Nuttall} window. Hence, using a combination of extended {\it Blackman--Nuttall} band shape with certain members of the family of {\it EoR window} ($1 \lesssim e \lesssim 2$) does not appear to improve MWA sensitivity, but offers a significant leverage in reducing foreground contamination from the power spectrum, thereby providing a cleaner {\it EoR window}. This could become very significant when imperfect source subtraction (position and calibration errors) and extended emission from extragalactic and Galactic foregrounds are also taken into account. This is due to the 7--8 orders of magnitude of extra tolerance provided by an extended {\it Blackman--Nuttall} window relative to a rectangular window in the amount of spillover of foreground contamination into the {\it EoR window}.

\begin{figure*}[htb]
\centering
\subfloat[][Sensitivity for $e=0$]{\label{fig:SNR1D_6hrs_8_multi_e0}\includegraphics[width=0.49\linewidth]{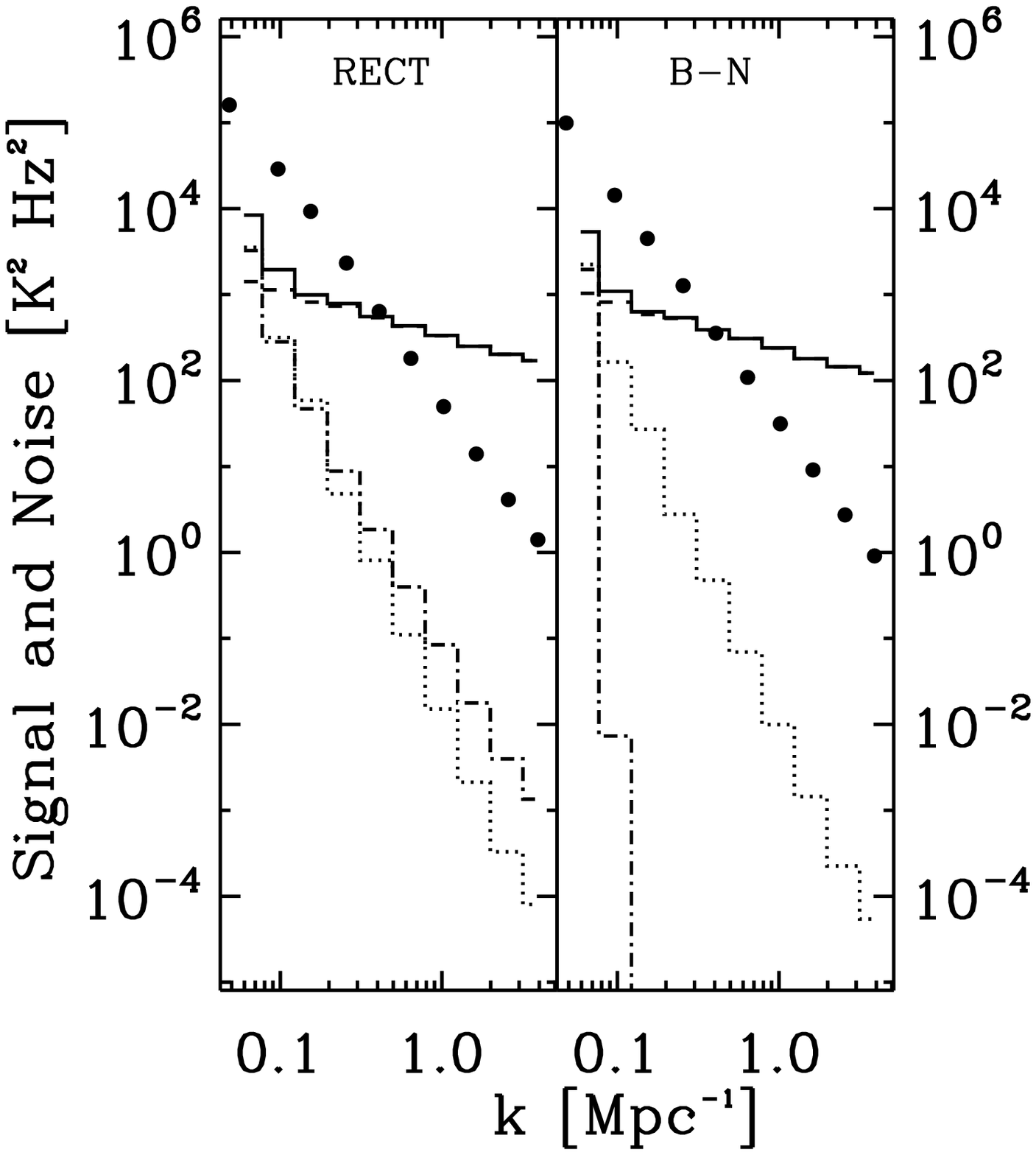}}
\subfloat[][S/N ratio]{\label{fig:case4:SNR}\includegraphics[width=0.49\linewidth]{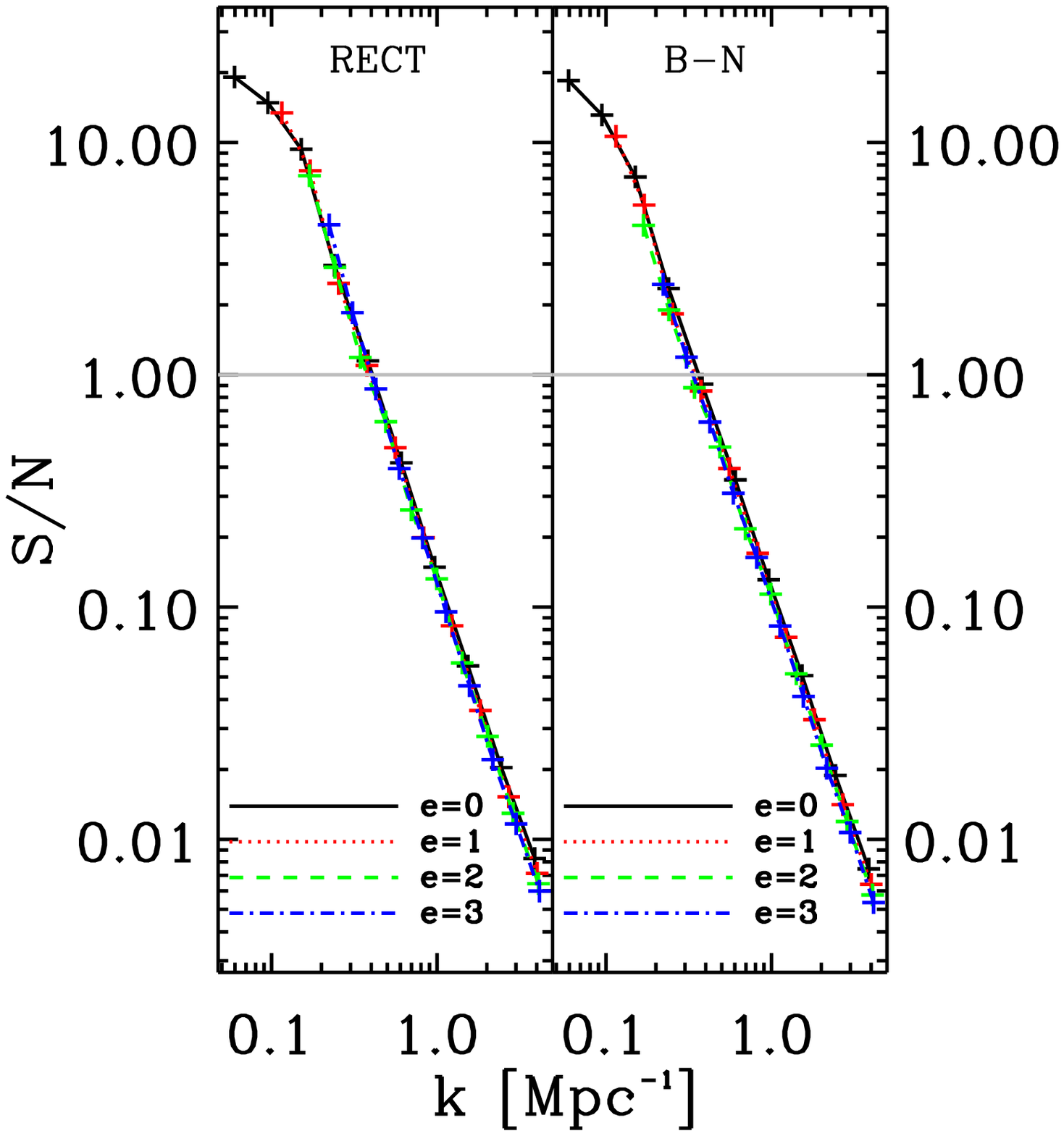}}
\caption{Properties of signal and uncertainty components in power spectrum (in units of K$^2$~Hz$^2$) for observing mode~(2), similar to Figure~\ref{fig:SN_EoR_windows_case_3}. The left and right sub-panels in each panel are obtained with rectangular and {\it Blackman--Nuttall} windows, respectively. ({\it a})~Averaged signal (solid circles) and uncertainties in different line styles (foregrounds: dot-dashed, thermal noise: dashed, sample variance: dotted, and total uncertainty: solid) in spherical shells of $k$ for $e=0$. ({\it b})~S/N ratios for different values of $e$ (shown in legend). `+' symbols denote the mean radii of spherical shells. EoR H{\sc i} power spectrum is detectable (S/N~$>1$) for $k\lesssim 0.4$~Mpc$^{-1}$.}
\label{fig:SN_EoR_windows_case_4}
\end{figure*}

We investigate the sensitivity for observing mode~(2), where a total of 1000~hours were divided over 20 patches of sky to obtain 20 independent measurements of power spectrum. This differs from observing mode~(1) in that the total observing time is now divided on multiple fields. Figure~\ref{fig:SNR1D_6hrs_8_multi_e0} is the counterpart of Figure~\ref{fig:SNR1D_6hrs_167_single_e0} and illustrates the signal and different uncertainty components obtained with the MWA in this observing mode. Counterparts to Figures~\ref{fig:case3:norm_EoR_FG_PS} and \ref{fig:case3:norm_TN_PS} will be identical and are not shown. Figure~\ref{fig:case4:SNR} shows the ratio of signal to total uncertainty (S/N) for this observing mode. 

Since the time on individual fields has reduced by a factor of $N_\textrm{fields}=20$ leading to a reduction in the number of coherent visibility measurements per field, the thermal noise component in each measurement of the power spectrum has increased by the same factor. The foreground contamination and sample variance in an individual power spectrum measurement remain identical to those in observing mode~(1) where only a single field is observed. When independent measurements of power spectra are averaged, all the components of uncertainty in the averaged power spectrum are reduced by a factor $\sqrt{N_\textrm{fields}}\approx 4.47$ relative to what they were in the individual measurements of power spectra. The net result, relative to observing mode~(1), is that the foreground contamination and sample variance have reduced by a factor 4.47, while thermal noise component has worsened by the same amount. This is evident when Figure~\ref{fig:SNR1D_6hrs_8_multi_e0} is compared with Figure~\ref{fig:SNR1D_6hrs_167_single_e0}. As a result, the crossover, up to which foreground contamination and sample variance dominate over thermal noise, moves leftward and now occurs at $k\simeq 0.1$~Mpc$^{-1}$ for both bandpass windows. This has a two-fold effect on overall sensitivity relative to that in the previous case: (a)~the sensitivity in lowest bin of $k$ ($k\simeq 0.06$~Mpc$^{-1}$), where sample variance and foreground components dominated over thermal noise component in observing mode~(1) has improved to 20, a factor of $\approx 4$ (consistent with $\sqrt{N_\textrm{fields}}\approx 4.47$); and, (b)~the sensitivity in other bins of $k$ ($k\gtrsim 0.1$~Mpc$^{-1}$) has degraded because the thermal noise component, which was already dominant in this regime, has worsened. The final effect on detectability is that S/N~$>1$ only for $k\lesssim 0.4$~Mpc$^{-1}$. 

How does sensitivity in observing mode~(2) compare to that in observing mode~(1)? Sensitivity is the result of a complex interplay between the relative magnitudes of the desired EoR H{\sc i} signal and different uncertainty components in the power spectrum. As far as MWA is concerned, thermal noise component appears to be dominant on scales with $k\gtrsim 0.1$--$0.2$~Mpc$^{-1}$. Hence, dividing the observing time over multiple fields and averaging the independent measurements of power spectra helps improve the sensitivity by a factor of $\approx 4$ for $k\lesssim 0.1$~Mpc$^{-1}$ but degrades it everywhere else. Consequently, the zone of detectability (S/N~$>1$) in $k$--space becomes narrower from $k\lesssim 0.8$~Mpc$^{-1}$ to $k\lesssim 0.4$~Mpc$^{-1}$. Thus, except for an improvement in sensitivity on the largest scales ($k\lesssim 0.1$~Mpc$^{-1}$), increasing the number of independent fields seems to offer no significant advantage. 

As far as effects of bandpass windowing and family of {\it EoR windows} are concerned, significant improvement in MWA sensitivity is neither seen with bandpass window shapes nor with the {\it EoR window} parameter $e$. However, it is crucial to obtain the cleanest {\it EoR window} possible in order to reduce the amount of systematics in the data, which may not be fully understood, such as those arising from unsubtracted foreground residuals. We find that refinements to the {\it EoR window} (through parameter $e$), a {\it Blackman--Nuttall} bandpass window shape and a combination of both significantly reduce the extragalactic foreground contamination in the measured power spectrum. 

\section{Summary}\label{sec:summary}

The primary goal of this work was to understand and estimate some of the fundamental factors that limit sensitivity of an EoR H{\sc i} power spectrum measurement, namely, point--like extragalactic foreground contamination, thermal noise and sample variance of the H{\sc i} brightness temperature fluctuations. A secondary goal was to understand how these uncertainties compete with each other on different scales in determining the sensitivity of a radio interferometer array, such as the MWA, towards detection of EoR H{\sc i} power spectrum. 

An analytic-{\it cum}-statistical approach was used in representing residual image cubes by assuming a radio source count distribution on the sky. For the 128--tile MWA at 170.7~MHz, when sources above the $5\,\sigma$ classical source confusion threshold were subtracted, the $1\sigma$ classical source confusion limit near the zenith was found to be $\approx 35$~mJy for a natural weighting scheme. 

Unsubtracted foreground sources and their sidelobes contaminate the predicted EoR signal. The frequency dependence of synthesized beam distributes the contamination from sidelobes onto a wedge-shaped region in $k$--space. We have presented a unified framework for signal and noise estimation using which we estimate foreground contamination in three--dimensional $k$--space. This framework also attempts to take into account {\it multi--baseline mode--mixing} effects caused by loss of coherence between non-identical baselines inside an independent cell in the spatial frequency domain. Using this framework, we establish an expression for the boundary of the wedge set by the {\it horizon limit}. 

We show for the first time, quantitatively, how the usage of a finite bandpass spills the contamination from unsubtracted sources and their sidelobes into the {\it EoR window}. This spillover decreases by 7--8 orders of magnitude, for instance, in the range $0.2$~Mpc$^{-1}\lesssim k_\parallel\lesssim 5$~Mpc$^{-1}$ at $k_\perp\approx 0.01$~Mpc$^{-1}$, by switching from a rectangular to an extended {\it Blackman--Nuttall} window. We argue this additional tolerance provided by the latter could prove to be of crucial significance in minimizing power spectrum contamination when the impact of imperfect source subtraction (due to position and calibration errors), and extended extragalactic and Galactic foregrounds are also considered. The frequency weighting in an extended {\it Blackman--Nuttall} bandpass window also lowers the thermal noise component of the power spectrum by 28\% relative to that achievable with a rectangular window. Conversely, in order to achieve the same thermal noise power with both windows, the duration of observing with a rectangular window has to be $\approx 40$\% longer when compared to an extended {\it Blackman--Nuttall} window.

We performed case studies of two different observing modes -- 6~hour synthesis repeated for a total of 1000~hours on a single field, and 6~hour synthesis repeated on 20 independent fields for a total of 1000~hours -- and studied the effects of the aforementioned uncertainties on EoR H{\sc i} power spectrum detection using the MWA. In both cases, detection appears to be possible (S/N~$>1$). 1000~hours on a single field shows the signal is detectable on scales with $k\lesssim 0.8$~Mpc$^{-1}$, while dividing it over 20 independent fields narrows the zone of detectability to scales with $k\lesssim 0.4$~Mpc$^{-1}$. Since sample variance and foregrounds, rather than thermal noise, are the dominant uncertainties on the largest scales ($k\lesssim 0.1$~Mpc$^{-1}$), detection sensitivity on these scales improves by roughly 4 times if the observing time is divided over 20 independent regions of sky. 

The concept of {\it EoR window} was probed quantitatively. Foreground contamination can be drastically reduced by using an extended {\it Blackman--Nuttall} bandpass window and through refinements ($1\lesssim e\lesssim 2$) to the {\it EoR window}.  

By modeling in detail the various uncertainties, we have shown the significance of different uncertainties on various scales and their roles in determining overall sensitivity. Observing many independent fields of view worsens the thermal noise and hence degrades the sensitivity for the MWA relative to a single field observation of the same duration. Bandpass window shaping and refinements to the {\it EoR window} do not affect sensitivity, but have a significant effect on containing the foreground contamination. \par\smallskip

\acknowledgments

This scientific work makes use of the Murchison Radio-astronomy Observatory. We acknowledge the Wajarri Yamatji people as the traditional owners of the Observatory site. Support for the MWA comes from the U.S. National Science Foundation (grants AST-0457585, PHY-0835713, CAREER-0847753, and AST-0908884), the Australian Research Council (LIEF grants LE0775621 and LE0882938), the U.S. Air Force Office of Scientific Research (grant FA9550-0510247), and the Centre for All-sky Astrophysics (an Australian Research Council Centre of Excellence funded by grant CE110001020). Support is also provided by the Smithsonian Astrophysical Observatory, the MIT School of Science, the Raman Research Institute, the Australian National University, and the Victoria University of Wellington (via grant MED-E1799 from the New Zealand Ministry of Economic Development and an IBM Shared University Research Grant). The Australian Federal government provides additional support via the National Collaborative Research Infrastructure Strategy, Education Investment Fund, and the Australia India Strategic Research Fund, and Astronomy Australia Limited, under contract to Curtin University. We acknowledge the iVEC Petabyte Data Store, the Initiative in Innovative Computing and the CUDA Center for Excellence sponsored by NVIDIA at Harvard University, and the International Centre for Radio Astronomy Research (ICRAR), a Joint Venture of Curtin University and The University of Western Australia, funded by the Western Australian State government. The National Radio Astronomy Observatory is a facility of the National Science Foundation operated under cooperative agreement by Associated Universities, Inc.

\appendix

\section{Estimating Classical Source Confusion Noise}\label{app:confusion_theory}

At any given frequency, the field of view contains many sources. Any given synthesized beam area of an array consists of many unresolved sources along that line of sight. Due to the statistical nature of the distribution of sources, the flux density contained in a synthesized beam area varies across the sky. The classical source confusion is the variation in flux density due to random distribution of unresolved sources across different beam areas on the sky. The theory on confusion is discussed in detail in \citet{con74} and \citet{roh00}. This paper uses the discussion and notations presented in the latter. 

The differential number density of sources per unit solid angle with respect to flux density is denoted by $\dif n/\dif S$, where $\dif n$ is the number of sources per steradian in a flux density interval between $S$ and $S+\dif S$. The variance in confusing source flux density in a given solid angle, $\Omega$, due to the number count distribution of sources is given by,
\begin{equation}\label{eqn:app:src_conf}
  \sigma_\textrm{C}^2=\Omega \int\limits_{S_\textrm{min}}^{S_\textrm{C}} S^2\,\frac{\dif n}{\dif S}\,\dif S,
\end{equation} 
where $S_\textrm{min}$ and $S_\textrm{C}$ are the lower and upper limits on the flux density of radio sources respectively. 

Since $\sigma_\textrm{C}^2$ is determined from the entire range of flux densities up to $S_\textrm{C}$, if there are bright sources, $\sigma_\textrm{C}^2$ will be overestimated. Hence, we perform an iterative procedure wherein all foreground sources brighter than $\rho_\textrm{C}\,\sigma_\textrm{C}$ and their sidelobes are subtracted, thereby updating the upper end of the flux density range, $S_\textrm{C}$. Here, $\rho_\textrm{C}$ acts as the source subtraction threshold factor. This iterative procedure of subtracting foreground sources brighter than $S_\textrm{C}=\rho_\textrm{C}\,\sigma_\textrm{C}$, and updating $S_\textrm{C}$ and $\sigma_\textrm{C}$, is performed until there are no sources brighter than $\rho_\textrm{C}\,\sigma_\textrm{C}$. We consider the $\sigma_\textrm{C}^2$ so computed as the ``true'' classical source confusion variance. 

In practice, usually, deconvolution procedures can estimate and subtract foreground sources and their associated sidelobes to a limited extent leaving behind a residual image. With the knowledge of distribution of radio sources encoded in $\dif n/\dif S$, a set threshold factor ($\rho_\textrm{C}$), and the solid angle ($\Omega$) corresponding to the angular resolution limit, the depth of source subtraction and corresponding residuals in the residual image can be determined using the iterative procedure described above, and is given by the following equation \citep{roh00}: 
\begin{equation}\label{eqn:app:src_conf_recursive}
  \rho_\textrm{C}^2 = \frac{S_\textrm{C}^2}{\Omega\displaystyle\int\limits_{S_\textrm{min}}^{S_\textrm{C}} S^2\,\frac{\dif n}{\dif S}\,\dif S}.
\end{equation}
This is obtained by imposing the criterion that the residual image is allowed to have unsubtracted sources of flux densities up to $S_\textrm{C}\leq \rho_\textrm{C}\,\sigma_\textrm{C}$. We have assumed that the source subtraction threshold, $S_\textrm{C}$, is set only by the classical source confusion noise whereas, in practice, confusion caused by sidelobes in a residual image will also play a role in determining the source subtraction threshold in this iterative procedure.

We use the radio source statistics provided by \citet{hop03}. Their best-fit expression for the source counts is:
\begin{equation}\label{eqn:app:logn-logS}
  \log\bigl[(\dif n/\dif S)/(S^{-2.5})\bigr] = \sum_{i=0}^6 a_i \,\bigl[\log (S/\textrm{mJy})\bigr]^i,
\end{equation}
valid at 1400~MHz for $0.05\,$mJy$\,\leq S \leq 1000\,$mJy, where $a_0=0.859$, $a_1=0.508$, $a_2=0.376$, $a_3=-0.049$, $a_4=-0.121$, $a_5=0.057$, and $a_6=-0.008$. $S$ is in units of mJy, and $(\dif n/\dif S)/S^{-2.5}$ is in units of Jy$^{1.5}$~sr$^{-1}$. The normalization by $S^{-2.5}$ indicates the expression is relative to a Euclidean universe. 

\begin{figure}[htb]
\centering
\subfloat[][modified differential source counts]{\label{fig:logn-logS}\includegraphics[width=0.49\linewidth]{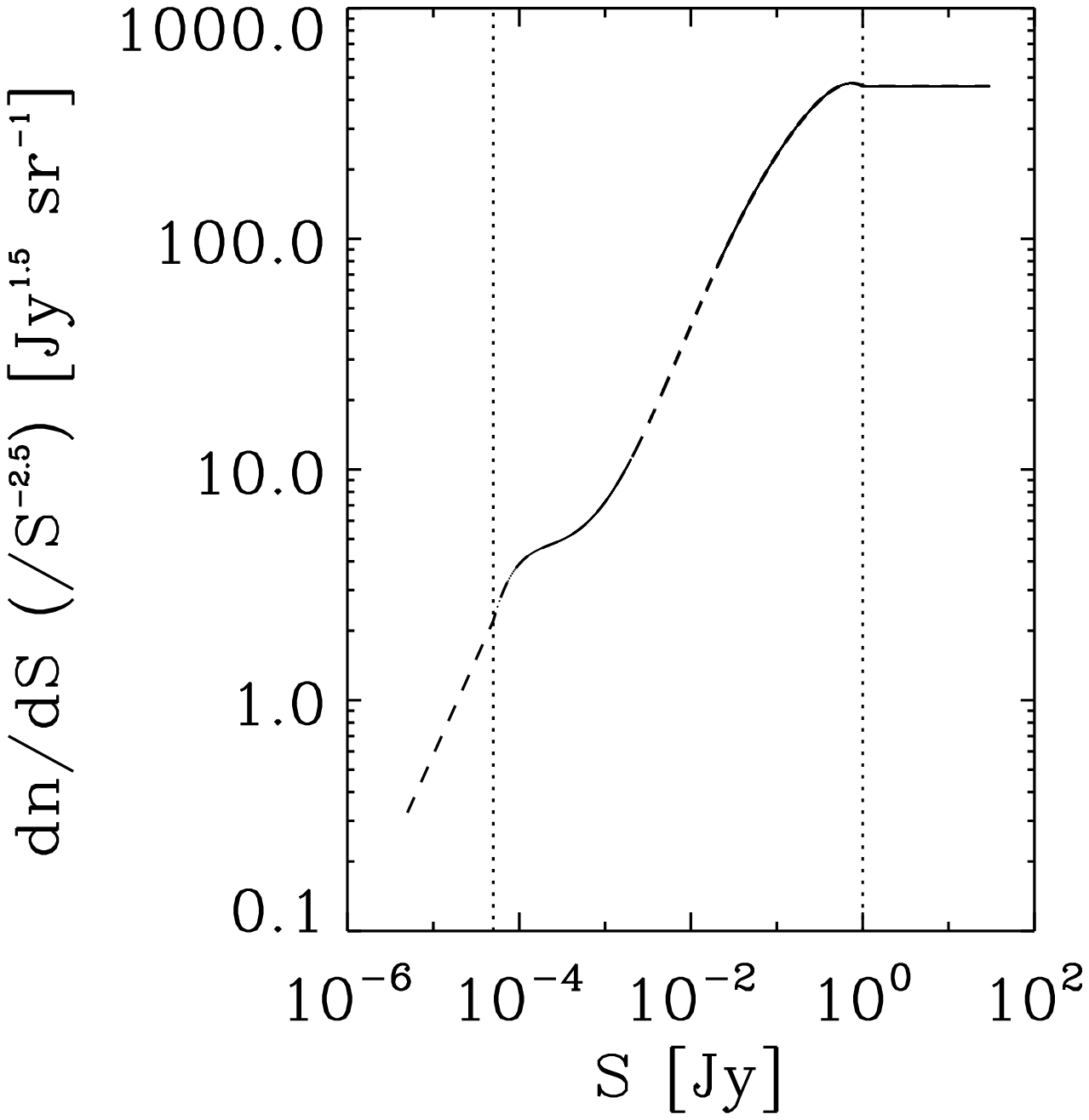}}
\subfloat[][confusion noise vs. solid angle]{\label{fig:src_conf_solid_angle}\includegraphics[width=0.49\linewidth]{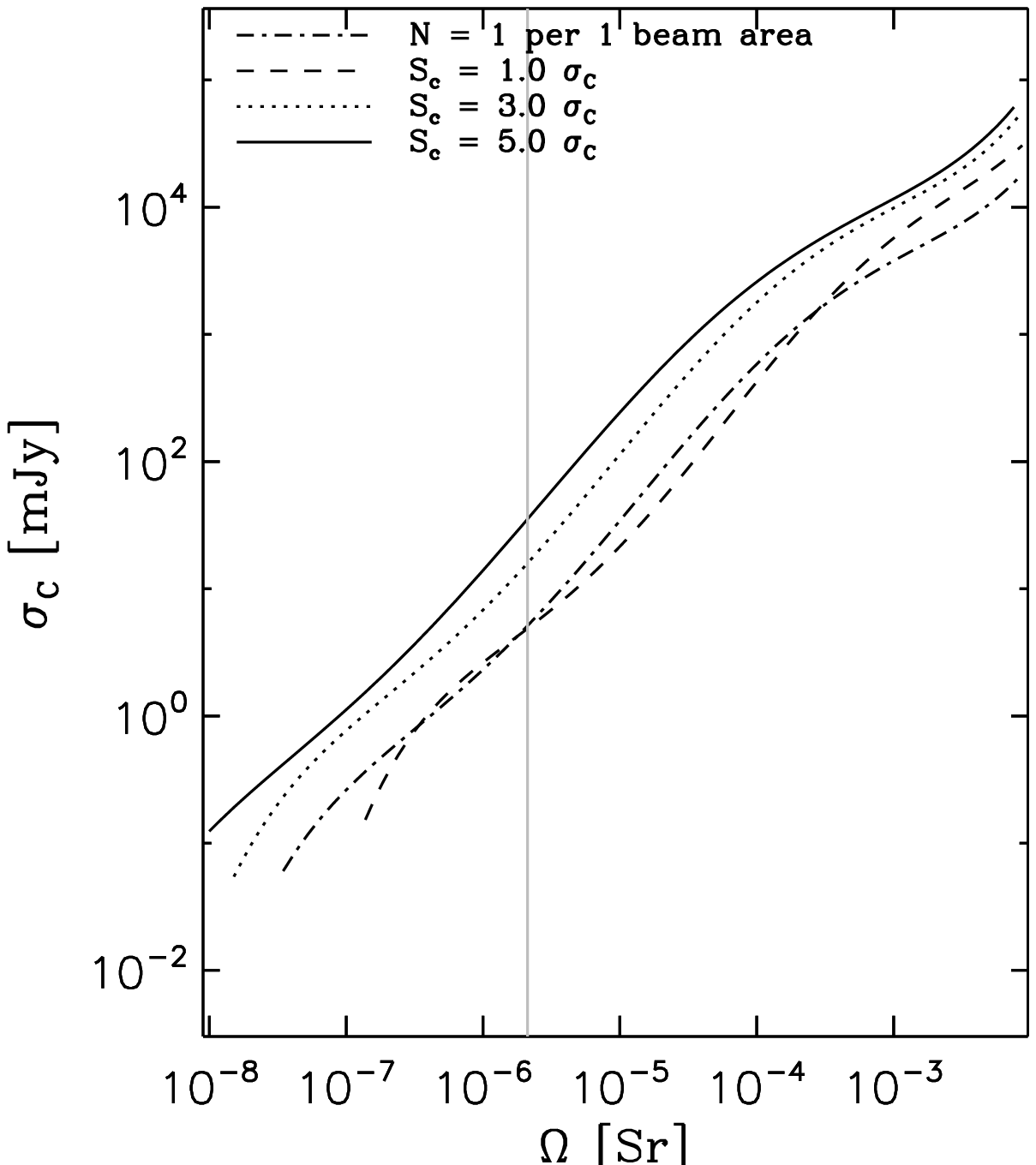}}
\caption{{\it Left}: Modified source count distribution of \citet{hop03} from equation~(\ref{eqn:app:logn-logS}) extrapolated outside the range 0.05~mJy--1~Jy (shown as vertical dotted lines) at 1400~MHz. The flat extrapolation for $S>1$~Jy is consistent with a Euclidean geometry representative of the local universe. The extrapolation for $S<0.05$~mJy has an identical slope to that in the flux density range 2--20~mJy (dashed lines). {\it Right}: The dependence of classical source confusion ($1\sigma$) on solid angle at 170.7~MHz for various values of thresholds, $\rho_\textrm{C}$, as indicated in the legend. This is derived from the extrapolated version of equation~(\ref{eqn:app:logn-logS}) shown in Figure~\ref{fig:logn-logS}. $\sigma_\textrm{C}$ depends on the choice of $\rho_\textrm{C}$: it increases with $\rho_\textrm{C}$. The dash-dotted line denotes the cutoff $S_\textrm{C}$ at which on average one source of flux density exceeding the cutoff is expected \citep{sub02}. The gray vertical line at $\Omega\approx 2\times 10^{-6}$~sr denotes the typical size of a resolution element for a naturally weighted image obtained with 128--tile MWA.}
\label{fig:FG_stats}
\end{figure}

The solid angle at various locations of an image in ($l,m$)--coordinates scales as:
\begin{align}\label{eqn:app:solid_angle}
  \Omega &= \frac{\Delta l\,\Delta m}{\sqrt{1-l^2-m^2}},
\end{align}
where $\Delta l\,\Delta m$ is the pixel size. This implies, from equation~(\ref{eqn:app:src_conf_recursive}), that confusion variance ($\sigma_\textrm{C}$), and flux density cutoff ($S_\textrm{C}$) are functions of position in the residual image. Estimating confusion noise over a wide range of solid angles requires an extrapolation of the empirical function in equation~(\ref{eqn:app:logn-logS}) on both ends of the flux density range. We have extrapolated at the higher end by a flat function mimicking a Euclidean local universe behavior, and at the lower end by an extension of the same slope found between 2--20~mJy. Figure~\ref{fig:logn-logS} shows equation~(\ref{eqn:app:logn-logS}) extrapolated at both ends beyond the aforementioned range in flux density (specified by vertical dotted lines). The dashed segments $S<0.05$~mJy and 2~mJy$\leq S\leq$20~mJy have slopes identical to each other. 

Under the assumption that the source population remains the same at the relevant frequency (170.7~MHz, for instance) under consideration as at 1400 MHz, and that these sources are unresolved with the MWA, the same expression for the distribution of source counts can be used once the spectral index is taken into account. There has been conflicting evidence in literature \citep[][and references therein]{ran12} over the spectral index properties of radio sources at frequencies below 1.4~GHz and whether the spectral index flattens for faint sources at low frequencies. Further, \citet{kel64} has pointed out that spectral index distribution is not independent of the observing frequency since sources with flatter spectral index are more likely to be observed at higher frequencies and those at lower frequencies tend to have steeper spectral index, and has subsequently provided a spectral index correction that tends to offset this bias. However, for our work, we have adopted a mean spectral index for radio sources as $\alpha = -0.78$ \citep{ish10}, where $S\,\propto f^{\alpha}$. Compared to flatter values of spectral index, our adopted value of $\alpha=-0.78$ gives us higher values for the classical source confusion at 170.7~MHz, making our sensitivity estimates conservative.

Sources can be subtracted down to various levels of threshold factor, $\rho_\textrm{C}$. Figure~\ref{fig:src_conf_solid_angle} represents a numerical solution for equation~(\ref{eqn:app:src_conf_recursive}) using $dn/dS$ from equation~(\ref{eqn:app:logn-logS}) for various values of $\rho_\textrm{C}$, as a function of the beam solid angle, $\Omega$. As expected from equation~(\ref{eqn:app:src_conf_recursive}), Figure~\ref{fig:src_conf_solid_angle} shows that $\sigma_\textrm{C}$ and $S_\textrm{C}$ are non-linear functions of both $\rho_\textrm{C}$ and $\Omega$. $\sigma_\textrm{C}$ depends on the choice of $\rho_\textrm{C}$: it increases with $\rho_\textrm{C}$. 

\section{Foreground contamination}\label{app:FG}

Following the notations established in \S\ref{sec:measurements}, the true visibilities of foregrounds, $V_{uvf}^\textrm{FG;T}(\underline{\mathbf{u}},f)$, are modified by the instrument as:
\begin{align}\label{eqn:FG_vis_gen}
  V_{uv\eta}^\textrm{FG}(\overline{\mathbf{u}}) &= \int V_{uvf}^\textrm{FG;T}(\underline{\mathbf{u}},f)\,S_{uv}(\underline{\mathbf{u}})\,W_f^\textrm{B}(f) \ast W_{uv}^\textrm{P}(\underline{\mathbf{u}})\,e^{-j2\pi\eta f}\dif f \notag\\
  &= V_{uv\eta}^\textrm{FG;T}(\overline{\mathbf{u}})\,S_{uv}(\underline{\mathbf{u}}) \ast W_\eta^\textrm{B}(\eta) \ast W_{uv}^\textrm{P}(\underline{\mathbf{u}}) \notag\\
  &= V_{uv\eta}^\textrm{FG;T}(\overline{\mathbf{u}})\,S_{uv}(\underline{\mathbf{u}}) \ast W_{uv\eta}(\overline{\mathbf{u}}).
\end{align}
where the true foreground visibilities, $V_{uvf}^\textrm{FG;T}(\underline{\mathbf{u}},f)$, have been convolved with the spatial frequency response of the antenna's power pattern, $W_{uv}^\textrm{P}(\underline{\mathbf{u}})$, multiplied by the sampling function, $S_{uv}(\underline{\mathbf{u}})$, in the ($u,v$)--plane, multiplied by the bandpass window function in frequency, $W_f^\textrm{B}(f)$, and Fourier transformed along frequency to obtain the measurements in Fourier space ($\overline{\mathbf{u}}$). $W_\eta^\textrm{B}(\eta)$ is the Fourier transform of $W_f^\textrm{B}(f)$, and $j=\sqrt{-1}$. $W_{uv\eta}(\overline{\mathbf{u}})=W_{uv\eta}^\textrm{P}(\overline{\mathbf{u}})\ast W_\eta^\textrm{B}(\eta)$, where $W_{uv\eta}^\textrm{P}(\overline{\mathbf{u}})$ may be interpreted as the spatial frequency response of the antenna's power pattern over an infinitely uniform bandpass. Assuming changes in the antenna power pattern over the observing band are insignificant, $W_{uv}^\textrm{P}(\underline{\mathbf{u}})=W_{uv\eta}^\textrm{P}(\underline{\mathbf{u}},\eta=0)$.

The covariance matrix for the measured foreground visibilities in Fourier space may be written as:
\begin{align}
  C^\textrm{FG}(\overline{\mathbf{k}}_i,\overline{\mathbf{k}}_j) &= \Bigl\langle V_{uv\eta}^\textrm{FG}(\overline{\mathbf{u}}_i)^\star V_{uv\eta}^\textrm{FG}(\overline{\mathbf{u}}_j)\Bigr\rangle \notag\\
  &= \iiint\,\iiint \Bigl\langle V_{uv\eta}^\textrm{FG;T}(\overline{\mathbf{u}}_p)^\star V_{uv\eta}^\textrm{FG;T}(\overline{\mathbf{u}}_q)\Bigr\rangle\,S_{uv}(\underline{\mathbf{u}}_p)^\star\,S_{uv}(\underline{\mathbf{u}}_q)\,W_{uv\eta}(\overline{\mathbf{u}}_i-\overline{\mathbf{u}}_p)^\star W_{uv\eta}(\overline{\mathbf{u}}_j-\overline{\mathbf{u}}_q)\,\dif^3 \overline{\mathbf{u}}_p\, \dif^3 \overline{\mathbf{u}}_q,
\end{align}
where $\star$ denotes a complex conjugate. But, being an uncorrelated statistical signal, $\Bigl\langle V_{uv\eta}^\textrm{FG;T}(\overline{\mathbf{u}}_p)^\star V_{uv\eta}^\textrm{FG;T}(\overline{\mathbf{u}}_q)\Bigr\rangle = P^\textrm{FG}(\overline{\mathbf{u}}_p)\,\delta(\overline{\mathbf{u}}_p-\overline{\mathbf{u}}_q)$. Hence,
\begin{align}\label{eqn:cov_FG}
  C^\textrm{FG}(\overline{\mathbf{k}}_i,\overline{\mathbf{k}}_j) &= \iiint P^\textrm{FG}(\overline{\mathbf{u}})\,\left|S_{uv}(\underline{\mathbf{u}})\right|^2\,W_{uv\eta}(\overline{\mathbf{u}}_i-\overline{\mathbf{u}})^\star W_{uv\eta}(\overline{\mathbf{u}}_j-\overline{\mathbf{u}})\,\dif^3 \overline{\mathbf{u}} \notag\\
  &= \iiint P^\textrm{FG}(\overline{\mathbf{u}})\,\left|S_{uv}(\underline{\mathbf{u}})\right|^2\,W_{uv}^\textrm{P}(\underline{\mathbf{u}}_i-\underline{\mathbf{u}})^\star W_{uv}^\textrm{P}(\underline{\mathbf{u}}_j-\underline{\mathbf{u}})\,W_\eta^\textrm{B}(\eta_i-\eta)^\star\,W_\eta^\textrm{B}(\eta_j-\eta)\,\dif^3 \overline{\mathbf{u}}.
\end{align}

The power spectrum is simply the diagonal of the covariance matrix, i.e., when the intensities at locations are compared with themselves: $P^\textrm{FG}_\textrm{inst}(\overline{\mathbf{k}}) = C^\textrm{FG}(\overline{\mathbf{k}}_i,\overline{\mathbf{k}}_j)\,\delta_{ij}$. Thus,
\begin{align}\label{eqn:PS_FG}
  P^\textrm{FG}_\textrm{inst}(\overline{\mathbf{u}}) &= \iiint P^\textrm{FG}(\overline{\mathbf{u^\prime}})\,\left|S_{uv}(\underline{\mathbf{u^\prime}})\right|^2\,\left|W_{uv\eta}(\overline{\mathbf{u}}-\overline{\mathbf{u^\prime}})\right|^2\,\dif^3 \overline{\mathbf{u^\prime}}. 
\end{align}

An insight into {\it mode--mixing} may be obtained if the above expression for power spectrum is re-expressed as a Fourier transform of quantities in the $(l,m,f)$--coordinates as follows:
\begin{align}
  P^\textrm{FG}_\textrm{inst}(\overline{\mathbf{u}}) &= \iiint P_{lmf}^\textrm{FG}(\overline{\mathbf{l}})\,\ast\,\left|S_{lm}(\underline{\mathbf{l}})\right|^2\,\left|W_{lm}^\textrm{P}(\underline{\mathbf{l}})\right|^2\,\left|W_f^\textrm{B}(f)\right|^2\,e^{-j2\pi(\underline{\mathbf{u}}\cdot\underline{\mathbf{l}}+\eta f)}\,\dif^3 \overline{\mathbf{l}}.  
\end{align}
Confusion variance, $P_{lmf}^\textrm{FG}(\overline{\mathbf{l}})$, from extragalactic foreground sources will be considered as the cause for foreground contamination in the power spectrum. It may be computed using equations~(\ref{eqn:app:src_conf_recursive}) and (\ref{eqn:app:solid_angle}). When line-of-sight ($f$ and $\eta$) terms are dropped, the integrand in the above equation is consistent with that in equation~(24) of \citet{bow09}. 

We further assume that $P_{lmf}^\textrm{FG}$ is independent of frequency, provided the residuals have no spectral variations. Noting that $\underline{\mathbf{u}}=f\,\underline{\mathbf{x}}/c$, the above equation may be re-written as:
\begin{align}\label{eqn:PS_FG_mm}
  P^\textrm{FG}_\textrm{inst}(\overline{\mathbf{u}}) &= \iiint P_{lm}^\textrm{FG}(\underline{\mathbf{l}})\,\ast\,\left|S_{lm}(\underline{\mathbf{l}})\right|^2\,\left|W_{lm}^\textrm{P}(\underline{\mathbf{l}})\right|^2\,\left|W_f^\textrm{B}(f)\right|^2\,e^{-j2\pi\left(\frac{\underline{\mathbf{x}}\cdot\underline{\mathbf{l}}}{c}+\eta\right)f}\,\dif^2 \underline{\mathbf{l}}\;\dif f \notag\\
  &= \iint P_{lm}^\textrm{FG}(\underline{\mathbf{l}})\,\ast\,\left|S_{lm}(\underline{\mathbf{l}})\right|^2\,\left|W_{lm}^\textrm{P}(\underline{\mathbf{l}})\right|^2\,\ast\,\left|W_\eta^\textrm{B}(\eta)\right|^2\,\ast\,\delta\left(\eta+\frac{\underline{\mathbf{x}}\cdot\underline{\mathbf{l}}}{c}\right)\,\dif^2 \underline{\mathbf{l}},
\end{align} 
where $\underline{\mathbf{x}}$ is the baseline vector in units of distance. The argument of the {\it delta} function connects transverse spatial structure to that along the line-of-sight. This demonstrates the {\it mode--mixing} aspect of contamination from foregrounds in the power spectrum. 

\par\bigskip
\bibliographystyle{apj}
\bibliography{eor}

\end{document}